\documentclass[twocolumn]{aastex631}

\usepackage{graphicx}	
\usepackage{amsmath}	
\usepackage{soul}
\usepackage{enumerate}
\usepackage{bbm}

\newcommand{\dd}{\mbox{$\>\mathrm{d}$}}
\newcommand{\Lcal}{\mbox{$\mathcal{L}$}}

\newcommand{\kpc}{\>{\rm kpc}}
\newcommand{\kth}{\mbox{$k^\mathrm{th}$}}

\definecolor{ForestGreen}{RGB}{34,139,34}

\definecolor{kcolor}{RGB}{150,0,250}

\begin{document}

\title[Minimum-entropy constraints on galactic
potentials]{Minimum-entropy constraints on galactic potentials}

\author[0000-0002-0740-1507]{Leandro {Beraldo e Silva}}
\affiliation{Department of Astronomy and Astrophysics, University of Michigan, Ann Arbor, MI, USA}
\affiliation{Steward Observatory and Department of Astronomy,\\University of Arizona, 933 N. Cherry Ave., Tucson, AZ 85721, USA}
\affiliation{Observatório Nacional, Rio de Janeiro - RJ, 20921-400, Brasil}
\author[0000-0002-6257-2341]{Monica Valluri}
\affiliation{Department of Astronomy and Astrophysics, University of Michigan, Ann Arbor, MI, USA}
\author[0000-0002-5038-9267]{Eugene Vasiliev}
\affiliation{University of Surrey, Guildford, Surrey GU2 7XH, United Kingdom}
\author[0000-0001-6924-8862]{Kohei Hattori}
\affiliation{National Astronomical Observatory of Japan, 2-21-1 Osawa, Mitaka, Tokyo 181-8588, Japan}
\affiliation{The Institute of Statistical Mathematics, 10-3 Midoricho, Tachikawa, Tokyo 190-8562, Japan}
\affiliation{Department of Astronomy and Astrophysics, University of Michigan, Ann Arbor, MI, USA}
\author{Walter {de Siqueira Pedra}}
\affiliation{University of São Paulo, Institute of Mathematics and Computer Sciences, Av. Trab. São Carlense 400, 13566-590, São Carlos, SP, Brazil}
\affiliation{BCAM - Basque Center for Applied Mathematics, Mazarredo, 14. 48009 Bilbao, Spain}
\author[0000-0003-2594-8052]{Kathryne J. Daniel}
\affiliation{Steward Observatory and Department of Astronomy,\\University of Arizona, 933 N. Cherry Ave., Tucson, AZ 85721, USA}

\correspondingauthor{Leandro {Beraldo e Silva}}
\email{lberaldoesilva@on.br, lberaldoesilva@gmail.com}

\begin{abstract}
  A tracer sample in a gravitational potential, starting from a
  generic initial condition, phase-mixes towards a stationary
  state. This evolution is accompanied by an entropy increase, and the
  final state is characterized by a distribution function (DF) that
  depends only on integrals of motion (Jeans' theorem). We present a
  method to constrain a gravitational potential assuming a stationary
  (phase mixed) sample by minimizing the entropy the sample would have
  if it were allowed to phase-mix in trial potentials. This method
  avoids modeling the DF, and is applicable to any sets of
  integrals. We provide expressions for the entropy of DFs depending
  on energy, $f(E)$, energy and angular momentum, $f(E,L)$, or three
  actions, $f(\vec{J})$, and investigate the bias and statistical
  uncertainties in their estimates. We show that the method correctly
  recovers the parameters for spherical and axisymmetric
  potentials. We also present a methodology to characterize the
  posterior probability distribution of the parameters with an
  Approximate Bayesian Computation, indicating a pathway for
  application to observational data. Using $10^4$ tracers with
  $10\% (20\%)$-uncertainties in the 6D coordinates, we recover the
  flattening parameter $q$ of an axisymmetric potential with
  $\sigma_q/q\sim 5\% (10\%)$. The python module for the entropy
  estimators, \texttt{tropygal}, is made publicly available.
\end{abstract}


\section{Introduction}
\label{sec:intro}
The gravitational potential is a fundamental aspect of any galaxy,
determining its stellar orbits and, afterall, their observed light
distribution. In the Milky Way ({\bf MW}), we can measure 6D
coordinates for millions of stars with Gaia \citep[][]{Gaia} and
spectroscopic surveys such as APOGEE \citep[][]{Apogee}, LAMOST
\citep[][]{Lamost}, GALAH \citep[][]{Galah} and DESI-MWS
\citep[][]{Cooper2023}. With theoretical modeling, these data can be
translated into a detailed picture of the Galaxy's mass
distribution. Of particular interest is the MW's dark matter ({\bf
  DM}) halo shape, which may constrain different scenarios for its
composition \citep[e.g.][]{Valluri2022}. Since this component is not
directly observed, one needs to infer its mass distribution from
stars' positions and kinematics.

A non-exhaustive list of methods to recover the underlying potential
using a tracer sample includes: the virial theorem and its variants
\citep{Zwicky1933, Bahcall1981, Watkins2010}, Jeans modelling
\citep[e.g.][]{Rehemtulla2022}, the ``orbital roulette''
\citep{Beloborodov2004}, the marginalization over an arbitrary number
of distribution function ({\bf DF}) components \citep{Magorrian2014},
the generating-function method of \cite{Tremaine2018}, the
minimization of the entropy of tidal streams \citep{Penarrubia2012,
  Sanderson2015}, the ``orbital pdf'' method of \cite{Han2016,
  Li2024}, Orbital Torus Imaging \citep[][]{Price-Whelan2021} and the
Maximum-Likelihood DF fitting \citep[e.g.][]{McMillan2012,
  McMillan2013, Deason2021}.

In all these methods, further assumptions are required besides the
information in the observed dataset. For instance, for tracers
described by a DF, one needs to assume that they constitute a system
in dynamical equilibrium. Otherwise, any potential is consistent with
a DF describing a non-stationary system \citep{McMillan2012,
  Green2023}. As another example, when modelling tidal streams, the
equilibrium assumption is replaced by an equally strong one, that the
debris were initially localized in phase space.

From Jeans' theorem, the DF of a system in equilibrium can be written
as a function of integrals of motion only, reducing the 6D phase-space
to 3D or less \citep[][]{BT}. For instance, isotropic spherical
systems can be described by a DF $f = f(E)$, where $E$ is the star's
energy, while for anisotropic spherical systems we can assume
$f = f(E, L)$, where $L$ is the magnitude of the angular momentum. In
general, samples in realistic galactic potentials normally require
three integrals of motion. In practice, this dimension reduction is
fundamental for a more efficient use of data.

Assuming a DF that depends on less integrals than required (a
dimension reduction too severe) delivers incorrect results. In
contrast, assuming a DF depending on more integrals than required is
not the most efficient use of data since it does not reduce the
dimensions as much as possible. Adopting three integrals is a good
compromise between generality and efficiency.

Among all integrals, actions offer several advantages \citep[despite
the difficulties in estimating them in practice -- see
e.g.][]{Sanders2016}: the transformation from phase-space coordinates
$(\vec{r},\vec{v})$ to angle-action ones $(\vec{\theta}, \vec{J})$ is
canonical, thus ${\dd\vec{r}\dd\vec{v} = \dd\vec{\theta}\dd\vec{J}}$;
actions are adiabatic invariants, i.e. they are conserved under slow
changes in the potential; angles are restricted to $[0, 2\pi)$, and a
system in equilibrium (phase-mixed) is simply described by a
probability density function ({\bf pdf})\footnote{We reserve the term
  DF and the notation $f()$ to the probability density function in 6D,
  and the term pdf and notation $F()$ to probability density functions
  of integrals of motion.} in action space
$F(\vec{J}) = (2\pi)^3 f(\vec{J})$. With angle-action variables, the
Hamiltonian depends only on the momenta, $H = H(\vec{J})$, and the
angle-coordinates increase linearly with time
${\vec{\theta} = \vec{\Omega} t + \text{const}}$, where
${\vec{\Omega} = \partial H/\partial \vec{J}}$. The dynamics is
thereby reduced to that of ``free particles''.

In the action-based DF-fitting method developed by
\cite{McMillan2012,McMillan2013} and further applied and improved by
e.g. \cite{Ting2013,Trick2016,Hattori2021}, the tracer population is
assumed to be in equilibrium, and characterized by a DF
$f(\vec{J})$. The MW potential is constrained by fitting functional
forms for both the total potential and the tracer DF. If the potential
is the only function of interest, one further marginalizes over the DF
parameters. For instance, \cite{Hattori2021} adopt a model with 9
parameters for the potential and 7 parameters for the DF which are
later marginalized over, similarly to other works employing this
technique. A disadvantage of this method is that it assumes an
analytic expression for the DF, which in reality is unknown.

The main goal of the current paper is to improve on this aspect, by
not assuming any functional form for the DF -- for other methods with
this intent see e.g. \cite{Han2016}, \cite{Li2024} for spherically
symmetric potentials. This avoids the overhead of fitting the DF
parameters and possible biases introduced by the chosen
DF. Information on the DF is obtained through non-parametric entropy
estimates.

Consider a tracer sample in equilibrium, and described by an unknown
DF $f(\vec{r}, \vec{v})$. As for any DF, we can define the so-called
differential entropy as
\begin{equation}
 S[f] \equiv -\int f \ln f \, \dd^6\vec{w},
    \label{eq:S_def}
\end{equation}
where $\vec{w} = (\vec{r}, \vec{v})$. This entropy is invariant for
changes of variables, in particular to angle-action variables
evaluated in any potential. In the correct potential where the sample
is in equilibrium {and in the absence of geometric cuts or other
  selection effects, the DF
  $f(\vec{r}, \vec{v}) = f(\vec{\theta}, \vec{J})$ is uniform in
  $\vec{\theta}$, whose phase-space volume is $(2\pi)^3$. The entropy
  associated with the angles is then maximum, and to keep $S$
  invariant, that associated with the actions must be minimum. This
  can be easily shown if
  $f(\vec{\theta}, \vec{J}) = \mathcal{F}(\vec{\theta})F(\vec{J})$, in
  which case the entropy is just the sum of the entropies in action
  and angle spaces -- in particular, for the fully phase-mixed sample
  $\mathcal{F}(\vec{\theta}) = (2\pi)^{-3}$. In Appendix
  \ref{sec:appendix}, we show that a similar idea also applies to
  non-separable DFs, which can always be separated in terms of
  conditional pdfs,
  $f(\vec{\theta}, \vec{J}) =
  \mathcal{F}(\vec{\theta}|\vec{J})F(\vec{J})$. We then conclude that
  the correct potential is recovered by minimizing a quantity
  involving the entropy of the marginal pdf $F(\vec{J})$ -- see
  \cite{Magorrian2014}, sec. 5.2 for a simpler reasoning and an
  orbit-averaged interpretation.

  This quantity is actually the entropy of the future final 6D DF
  describing the sample if it were allowed to phase-mix in each trial
  potential. This final DF would be a different (and unknown) function
  of actions in each trial potential. Since actions are conserved, we
  estimate this final entropy right away for each potential, with no
  need to wait for phase-mixing, and the true potential is the one
  with minimum entropy. We also show that the same method is
  applicable to any sets of integrals, provided they respect the
  symmetry requirements of the problem. While one might try to fit
  potentials by instead maximizing an entropy in angle-space, in
  Appendix \ref{sec:max_entropy_angles} we discuss why this is not
  expected to work.

  Our approach is related to the minimum-entropy estimates of
  semi-parametric models \citep[][]{Wolsztynski2005}, where the
  potential is the parametric part and the pdf is the non-parametric
  one. In Sec.~\ref{sec:formalism} we describe the general formalism,
  starting from the action-based DF-fitting and showing how it is
  extended by our method. Sec.~\ref{sec:entropy} presents the
  expressions for the entropy estimator in the assumption-free (6D)
  case and in cases where the DF is a (unknown) function of integrals
  of motion. Sec.~\ref{sec:isochrone} shows the physical basis of the
  method, investigates the bias and variance of the entropy estimates
  for DFs depending only on integrals, and apply a bias correction. In
  Sec.~\ref{sec:min_entropy_illustra} we use a fixed sample that is
  phase-mixed in a given potential to illustrate that the entropy of
  the sample, estimated using integrals in different potentials, is
  minimum at the true potential. In Sec.~\ref{sec:fits} we demonstrate
  through actual fits that our method recovers the true parameters of
  a simple spherical potential, and of a flattened axisymmetric
  potential. We discuss our results in Sec.~\ref{sec:discussion} and
  summarize in Sec.~\ref{sec:summary}. The mathematical basis of the
  method is presented in Appendix \ref{sec:appendix}.

\section{General formalism}
\label{sec:formalism}
Assume a sample of $N$ stars in dynamical equilibrium in a
gravitational potential $\phi(\vec{r})$. Assume further that this is
an unbiased sample of an unknown underlying DF $f_0$ which, as allowed
by Jeans' theorem, is a function of integrals of motion in
$\phi(\vec{r})$ -- we focus on actions $\vec{J}$, but other integrals
can be used too. Our task is to use the 6D coordinates of these stars,
assume a functional form for $\phi(\vec{r})$ and constrain its
parameters.

To motivate the minimum-entropy method proposed in this work, we start
presenting the maximum-likelihood DF-fitting formalism. In the
DF-fitting method, one assumes functional forms for both the potential
$\phi(\vec{r})$ and for the DF $f(\vec{J}|\vec{p})$ describing the
tracer sample, where $\vec{p}$ encapsulates parameters of both the
potential and the DF. The DF is assumed to describe the stationary
state the given sample would achieve after phase-mixing in each trial
potential. For simplicity, we assume a full-sky sample in absence of
any selection function or observational errors -- the full treatment
is presented by e.g.: \cite{McMillan2013, Hattori2021}. In this case,
the likelihood for a star to have coordinates
$\vec{w}_i \equiv (\vec{r}_i, \vec{v}_i)$ is $f_i(\vec{J}_i|\vec{p})$,
where $\vec{J}(\vec{w}|\phi)$ are actions, which depend on the
potential, and $f(\vec{J}|\vec{p})$ is properly normalized. The sample
joint likelihood is ${\hat{\Lcal} = \prod_{i=1}^N f_i}$, and the
log-likelihood to be maximized is
\begin{equation}
\ln \hat{\Lcal}(\vec{w}|\vec{p}) = \sum_{i=1}^N \ln f_i(\vec{J}_i|\vec{p}),
\label{eq:lnL}
\end{equation}
with trial potentials entering the fit through the actions. Note that
Eq.~\eqref{eq:lnL} can be seen as an estimate\footnote{For any
  quantity $X$ we denote its estimate by $\hat{X}$.} of the ``true''
log-likelihood
\begin{equation}
    \ln \Lcal = -NH(f_0, f),
    \label{eq:lnL_expect}
\end{equation}
where 
\begin{equation}
    H(f_0, f) = - \int f_0 \ln f \, \dd\vec{w}
    \label{eq:cross_entropy_def}
\end{equation}
is the cross-entropy and $f_0 = f(\vec{J}|\vec{p}_0)$, with
$\vec{p}_0$ being the true parameters. Note that $H(f_0, f)$ is
minimum for $f = f_0$, illustrating that the likelihood is maximum at
the true parameters \citep[e.g.][]{Akaike1992}.

The formalism above concerns the DF-fitting method where an analytic
DF is assumed. It can be connected with the minimum-entropy method
presented here as follows. As before, we consider the stationary state
that the given sample would achieve after phase-mixing in each trial
potential. For each of these stationary states, from
Eq.~\eqref{eq:S_def}, the differential entropy of its DF can be
estimated via Monte Carlo with a sample of $f$ as
\begin{equation}
    \hat{S} = -\frac{1}{N} \sum_{i=1}^N \ln \hat{f}_i(\vec{J}_i|\vec{p}),
    \label{eq:S_hat}
\end{equation}
where $\hat{f}_i$ is an estimate of $f(\vec{J}_i|\vec{p})$, as
detailed in Sec.~\ref{sec:entropy}. Comparing Eqs.~\eqref{eq:lnL} and
\eqref{eq:S_hat} might suggest writing
\begin{equation}
     \ln \hat{\lambda}(\vec{p}) \equiv -N \hat{S}(\vec{p})
     \label{eq:lnlambda}
\end{equation}
for the ``log-likelihood''. However, despite appearances,
$\ln \hat{\lambda}(\vec{p})$ is not an estimate of the log-likelihood,
as can be seen by comparing Eq.~\eqref{eq:S_def} with
Eqs.~\eqref{eq:lnL_expect}-\eqref{eq:cross_entropy_def}. In other
words, a log-likelihood would involve assuming a functional form for
$f(\vec{J}|\vec{p})$ and estimating the cross-entropy between the true
DF $f_0=f(\vec{J}|\vec{p}_0)$ and trial DFs $f(\vec{J}|\vec{p})$. In
contrast, $\ln \lambda(\vec{p})$ involves estimating the entropy of
the (unknown) future DFs in trial potentials. Thus,
$\ln \lambda(\vec{p})$ only corresponds to the log-likelihood at the
best-fit model, i.e. $\ln \lambda(\vec{p_0}) = \ln
\Lcal(\vec{p_0})$. Besides, $\ln \hat{\lambda}$ is not a smooth
function of the parameters as required for a log-likelihood estimate,
but it is noisy since it is based on estimates of the DF, rather than
evaluating an analytical DF. However, as we demonstrate in practice in
Sec.~\ref{sec:fits}, and on mathematical grounds in Appendix
\ref{sec:appendix}, on average $\hat{S}$ has its minimum at
$\vec{p}_0$, and can be minimized to find the best-fit model -- see
Fig.~\ref{fig:lnL_illustra} for an illustration.

\begin{figure}
	\includegraphics[width=\columnwidth]{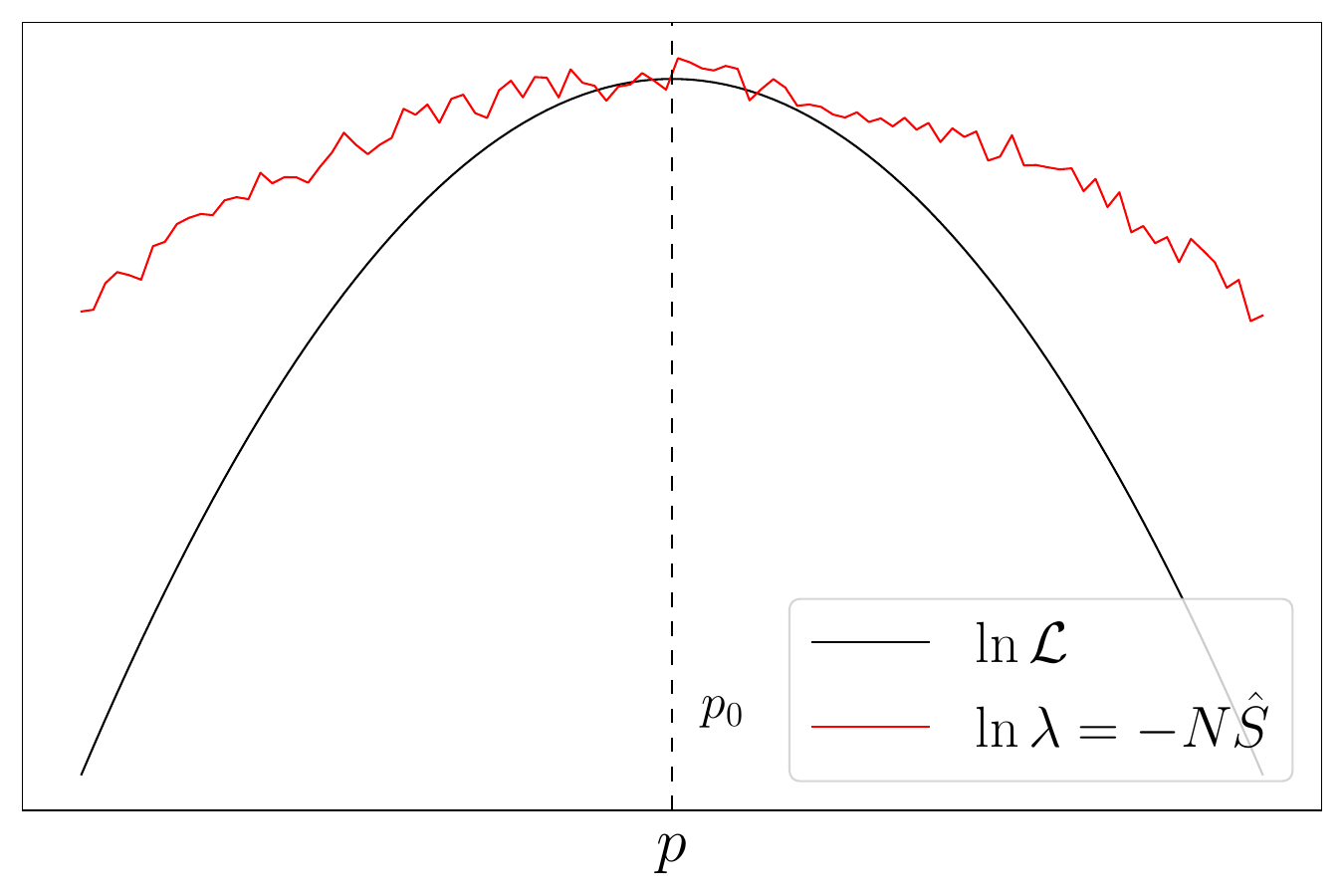}
        \caption{Illustrative comparison of the log-likelihood
          $\ln \mathcal{L}$ with the quantity used to find its
          maximum, $\ln \lambda$. Although being different quantities,
          on average they both peak at the same value $p_0$ and have
          the same value at the peak.}
    \label{fig:lnL_illustra}
\end{figure}

The maximum-likelihood principle is then replaced by a minimum-entropy
one, where we minimize the ``future entropy'' -- the entropy the
sample would reach after phase-mixing in each trial potential. The DF
describing these final states is always assumed to be a function of
integrals, in accordance with Jeans' theorem. However, we do not need
to assume any functional form for the DF, and in the remainder of this
work $\vec{p}$ encapsulates only parameters for the potential.

As illustrated in Fig.~\ref{fig:lnL_illustra}, fluctuations in
$\hat{S}$ can lead to misidentifying the best-fit model, and some
smoothing is required to avoid that. In this paper, we estimate the
entropy with the $k^\mathrm{th}$-Nearest-Neighbor method
\citep[k-NN,][]{Leonenko2008, Leonenko2008b}, and we smooth out
$\hat{S}$ by taking $k>1$ -- see
Sec.~\ref{sec:entropy_fluctuation}. After identifying the best-fit
model by minimizing the entropy of the DF, we perform an Approximate
Bayesian Computation to sample the posterior and get credible
intervals for the parameters (Sec.~\ref{sec:fits}). We remark that
$\ln \lambda$ was just introduced above to motivate our
minimum-entropy method with a conceptual link to the
maximum-likelihood principle, but in practice we simply minimize the
future entropy of the sample, with no further mention of
$\ln \lambda$.

Although in this paper we do not consider any selection effects or a
realistic survey footprint with geometric cuts, these are fundamental
aspects for the applicability of the method to real data. With real
data, we do not have a sample of the DF $f(\vec{w})$ assumed in
equilibrium. Rather, we have a sample of the DF
\begin{equation}
    f_S(\vec{w}) = \frac{f(\vec{w})\mathbb{S}(\vec{r})}{A},
\end{equation}
where $\mathbb{S}(\vec{r})$ is the selection function encapsulating
the footprint and spatial dependencies within it, and
${A = \int f(\vec{w})\mathbb{S}(\vec{r})\dd^6\vec{w}}$ is a
normalization constant. Replacing in Eq.~\eqref{eq:S_def}, we have
\begin{equation*}
 S = -\int \left(\frac{A}{\mathbb{S}(\vec{r})}f_S\right) \ln \left(\frac{A}{\mathbb{S}(\vec{r})}f_S\right) \, \dd^6\vec{w},
\end{equation*}
which is now a weighted differential entropy. This might make it
difficult to estimate the entropy $S$, since the original estimators
we discuss in Sec.~\ref{sec:entropy} are intended to use samples of
$f$. However, if the selection function $\mathbb{S}(\vec{r})$ is
known, the estimation method can be adapted to provide $S$ given
samples of $f_S$ -- see \cite{Ajgl2011}.

In this paper, we consider ideal full-sky samples with no selection
effects and set $A = \mathbb{S}(\vec{r}) = 1$. Having presented the
general formalism, we now present expressions to estimate the entropy
in general and in particular cases of DFs only depending on integrals
of motion.

\section{Entropy estimators}
\label{sec:entropy}
We start by defining the entropy of a DF $f(\vec{w})$. Instead of
Eq.~\eqref{eq:S_def}, we modify the entropy definition as
\begin{equation}
    S \equiv -\int f \ln \left(\frac{f}{\mu}\right) \, \dd^6\vec{w},
    \label{eq:S_def_2}
\end{equation}
where $\mu$ is such that the argument in $\ln (f/\mu)$ is
dimensionless, e.g. if
$[f] = \mathrm{length}^{-3}\,\mathrm{velocity}^{-3}$, it is convenient
to use coordinates normalized by their dispersions
$\sigma_{w_1},...,\sigma_{w_6}$, defining
${w_{1}'=w_1/\sigma_{w_1},\dots,w_{6}'=w_6/\sigma_{w_6}}$ and setting
${\mu=|\Sigma|^{-1}}$, where
$|\Sigma|=\sigma_{w_1}\dots\sigma_{w_6}$. With
$f'(\vec{w}') = |\Sigma| f(\vec{w})$, we have:
 \begin{equation}
    S= -\int f' \ln f' \, \dd^6\vec{w}' = -\int f \ln (|\Sigma|f) \, \dd^6\vec{w}.
    \label{eq:S_def_norm}
\end{equation}
For estimators with an isotropic kernel such as the k-NN discussed
below, this normalization works as to ``isotropize'' the coordinates,
whereas the entropy is made invariant by an appropriate change of
variables. From Eq.~\eqref{eq:S_def_norm},
$-\int f \ln f \, \dd^6\vec{w} = S + \ln |\Sigma|$. Another advantage
of the definition \eqref{eq:S_def_2} is that it allows us to
accommodate densities of states when using pdfs of integrals of
motion, as shown below.

Eq.~\eqref{eq:S_def_norm} is the invariant entropy we start from in
this section and from which we transform coordinates for the cases
where the DF is a function of integrals of motion only. For a sample
of $N$ points, it can be estimated as
\begin{equation}
    \hat{S} = -\frac{1}{N} \sum_{i=1}^N \ln \hat{f}_i',
    \label{eq:S_hat_norm}
\end{equation}
where $\hat{f}_i'$ is an estimate of $f'(\vec{w}_i')$. In principle,
any density estimator could be employed to estimate $f'(\vec{w}_i')$
-- see \cite{Silverman1986} for a review on density
estimates. However, for the particular purpose of estimating the
entropy with Eq.~\eqref{eq:S_hat_norm}, a few estimators have been
shown to be optimal \citep[see e.g.][]{Joe1989, Hall1993,
  Beirlant1997, Leonenko2008, Leonenko2008b} -- for a comparison of
different methods in $N$-body simulations, see \cite{BeS2017}. The
latter work demonstrated, in particular, a reasonable agreement of
entropy estimates based on k-NN and kernel density estimates, and the
high accuracy of the Fokker-Planck modelling of the collisional
relaxation, later confirmed on rigorous theoretical grounds by
\cite{Fouvry2021}. More recently, \cite{Modak2023} used this estimator
to study the eccentricity distribution of wide binaries.

Among the optimal methods, we use the k-NN estimator, which is fully
non-parametric and fast, since the neighbors' identification can be
optimized with kd-trees. This entropy estimator was introduced by
\cite{Kozachenko1987} for $k=1$ and later generalized for any $k$. In
this method, the plug-in density estimate is given by \citep[see
e.g.][and references therein]{Leonenko2008, Leonenko2008b, Biau2015,
  Berrett2019}:
\begin{equation}
    \hat{f}_i' = \frac{e^{\psi(k)}}{(N-1)V_d {D_{ik}}^d},
    \label{eq:f_est}
\end{equation}
and
\begin{equation}
    V_d = \pi^{d/2}/\Gamma(d/2+1)
    \label{eq:V_d}
\end{equation}
is the volume of the $d$-dimensional unit-radius hypersphere,
$D_{ik} = \sqrt{(\vec{r}_i' - \vec{r}_k')^2 + (\vec{v}_i' -
  \vec{v}_k')^2}$ is the Euclidean phase-space distance of particle
$i$ to its $\kth$ nearest neighbor, and $\psi(x)$ is the digamma
function\footnote{In particular, ${\psi(1) = -\gamma \approx -0.577}$
  (Euler-Mascheroni constant).}. For a sketch of a proof of
convergence of this method for $k=1$, see Appendix B of
\cite{Charzynska2015}.

Eq.~\eqref{eq:S_hat_norm} with Eq.~\eqref{eq:f_est} plugged in is a
proper entropy estimator in the sense that its bias and variance tend
to zero for $N\rightarrow\infty$. For the bias, the convergence speed
strongly depends on the dimension $d$ and regularity of $f$ \citep[see
sec. 7.4 and chap. 14 of][]{Biau2015}. Although the actual bias can
depend on particular features of the pdf, it is typically smaller in
lower dimensions, as we verify in Sec.~\ref{sec:entropy_bias}. The
expected variance scales as $\propto N^{-1}$, irrespective of the
dimension \citep[sec. 7.3 of][]{Biau2015}, as we verify in
Sec.~\ref{sec:entropy_fluctuation}.

Eq.~\eqref{eq:f_est} contrasts with naively estimating the density as
the number $k$ of points, besides point $i$, in the hypersphere around
point $i$, divided by its volume, which would introduce a
non-vanishing bias for $N\rightarrow \infty$. A slightly better
reasoning would provide better estimates, although yet not fully
bias-corrected: since the $k$-th neighbor is at the edge of the
hypersphere, a small volume around it is approximately half inside,
half outside the hypersphere, and it should count as ``half a
neighbor'' of $i$, estimating the pdf as
\begin{equation}
    \hat{\mathbbm{f}}_i = \frac{1}{N-1}\frac{k-1/2}{V_d {D_{ik}}^d}.
    \label{eq:f_est_2}
\end{equation}
The entropy estimate based on Eq.~\eqref{eq:f_est_2} differs from that
based on Eq.~\eqref{eq:f_est} by
\begin{equation*}
    \delta S = \ln (k-1/2) - \psi(k) = \ln (k-1/2) - \ln (k_\mathrm{eff} - 1/2),
\end{equation*}
where $k_\mathrm{eff} = e^{\psi(k)} + 1/2$ is an ``effective number of
nearest-neighbors''. For $k=1, 2, 3, 4$, it is, respectively,
$k_\mathrm{eff}\approx 1.06, 2.03, 3.02, 4.01$. For large $k$,
${e^{\psi(k)} \approx k -1/2 + \mathcal{O}(1/k)}$, and
$k_\mathrm{eff}\approx k$.

For two general distributions $f_0$ and $f$, we also re-define their
cross-entropy as
\begin{equation}
    H(f_0, f) \equiv -\int f_0 \ln \left(\frac{f}{\mu}\right) \, \dd^6\vec{w}.
    \label{eq:cross_entropy_def_2}
\end{equation}
Note that, in general, it is possible to estimate $H(f_0, f)$ even if
the samples of $f_0$ and $f$ have different sizes $N$ and $M$,
respectively. Eq.~\eqref{eq:cross_entropy_def_2} is estimated as
\begin{equation}
    \hat{H} = -\frac{1}{N} \sum_{i=1}^N \ln \hat{\xi}_i',
    \label{eq:H_hat_norm}
\end{equation}
where
\begin{equation}
    \hat{\xi}_i' = \frac{e^{\psi(k)}}{MV_d {D_{ik}}^d},
    \label{eq:xi_est}
\end{equation}
and $D_{ik}$ is the distance between point $i$ of the $f_0$-sample to
its $k$-nearest neighbor \emph{in the $f$-sample}
\citep[][]{Leonenko2008b}. We can interpret $\hat{\xi}_i'$ as an
estimate of $f$ at the point $i$ of the $f_0$-sample. In this paper,
we restrict to samples of equal sizes, so $M=N$, and normalize
coordinates by typical dispersions of the $f_0$-sample. To explore the
parameters' posterior distribution in Sec. \ref{sec:fit_isoc},
$f_0(\vec{J})$ will represent the (unknown) underlying DF describing
the sample in the best-fit potential and $f(\vec{J})$, the final
(equilibrium) DF of the sample in each trial potential.
 
Eqs.~\eqref{eq:S_hat_norm} and \eqref{eq:H_hat_norm}, with
Eqs.~\eqref{eq:f_est} and \eqref{eq:xi_est} respectively plugged in,
converge in probability to the true entropies under weak conditions on
the underlying DFs \citep[e.g.][]{Leonenko2008, Biau2015,
  Lombardi2016}. The python module \texttt{tropygal}\footnote{The
  documentation and installation instructions can be accessed at
  \url{https://tropygal.readthedocs.io/en/latest/}.} implements these
entropy estimators, as well as a few galactic dynamics models with
analytic DFs.

As explained in Sec.~\ref{sec:intro}, the method developed here
assumes the sample is phase-mixed in the true potential, and also
considers the entropy the sample would have if evolved until
phase-mixed in a trial potential. In the next subsections, we present
expressions for cases where the DF only depends on integrals of
motion, as required by Jeans' theorem for phase-mixed samples. In the
following, we denote $S_I= S[f(\vec{I})]$, i.e. the entropy of the DF
when $f$ is a function of integrals $\vec{I}$. Note that this differs
from the entropy of the integrals' pdf
$S[F(\vec{I})] = -\int F\ln F \dd\vec{I}$, as we show here and, in
more detail, in Appendix \ref{sec:appendix}.

\subsection{Isotropic spherical system, $f = f(E)$}
\label{sec:f_E}
For isotropic spherical systems in equilibrium, we can write
${f(\vec{w}) =f(E)}$, where $E = v^2/2 + \phi(r)$ and $\phi(r)$ is the
potential. In this case, Eq.~\eqref{eq:S_def_norm} reduces to
\begin{equation}
    S_\mathrm{E} = -\int F(E) \ln\left[ \frac{|\Sigma|F(E)}{g(E)}\right]\dd E,
    \label{eq:S_E_def}
\end{equation}
where 
\begin{equation}
    F(E) = f(E)g(E)
    \label{eq:F_E}
\end{equation}
is the pdf in energy space and
\begin{equation}
    g\left[E|\phi(r)\right] = (4\pi)^2 \int_0^{r_m(E)} r^2 \sqrt{2[E-\phi(r)]}\dd r
    \label{eq:g_E}
\end{equation}
is the density of states, with $r_m(E)$ being the radius where
$\phi = E$. If $\sigma_E$ is a typical energy dispersion, we define
$E'=E/\sigma_E$, and estimate $S_\mathrm{E}$, Eq.~\eqref{eq:S_E_def},
as

\begin{equation}
    \hat{S}_\mathrm{E} = -\frac{1}{N}\sum_{i=1}^N \ln \left[\frac{\hat{F_i}'(E_i')}{\mu(E_i)}\right],
    \label{eq:S_E_est}
\end{equation}
where $\mu(E) = \sigma_E |\Sigma|^{-1}g\left[E|\phi(r)\right]$. We
estimate $\hat{F_i}'(E_i')$, the energy pdf, with $d=1$ and
${D_{ik} = |E_i' - E_k'|}$ in Eq.~\eqref{eq:f_est}. If it is
convenient to write the density of states in terms of the normalized
energy and angular momentum, we can replace
$g\left[E|\phi(r)\right] = \sqrt{\sigma_E} g\left[E' \Big
  |\phi(r)/\sigma_E\right]$.

\subsection{Anisotropic spherical system, $f = f(E,L)$}
\label{sec:f_EL}
For anisotropic spherical systems with a DF ${f(\vec{w})=f(E, L)}$,
where $L = v_t r$ and $v_t^2=v_\theta^2 + v_\varphi^2$ in spherical
coordinates $(r,\theta,\varphi)$, Eq.~\eqref{eq:S_def_norm} reduces to
\begin{equation}
    S_\mathrm{EL} = -\int F(E,L) \ln\left[ \frac{|\Sigma|F(E,L)}{g(E,L)}\right]\dd E\dd L,
    \label{eq:S_EL_def}
\end{equation}
where the pdf for energy and angular momentum is 
\begin{equation}
    F(E,L) = f(E,L)g(E,L),
    \label{eq:F_EL}
\end{equation}
and the density of states is
\begin{equation}
    g\left[E,L|\phi(r)\right] = 8\pi^2 L T_r\left[E,L|\phi(r)\right].
    \label{eq:g_EL}
\end{equation}
 The period of radial motion $T_r\left[E,L|\phi(r)\right]$ is given by
\begin{equation}
    T_r\left[E,L|\phi(r)\right] = 2 \int_{r_\mathrm{per}}^{r_\mathrm{apo}} \frac{\dd r}{\sqrt{2[E-\phi(r)] - L^2/r^2}},
    \label{eq:T_r}
\end{equation}
with $r_\mathrm{per}$ and $r_\mathrm{apo}$ being the peri- and
apo-center distances. Defining $(E', L') = (E/\sigma_E, L/\sigma_L)$,
we estimate
\begin{equation}
    \hat{S}_\mathrm{EL} = -\frac{1}{N}\sum_{i=1}^N \ln \left[\frac{\hat{F_i}'(E_i', L_i')}{\mu(E_i, L_i)}\right],
    \label{eq:S_EL_est}
\end{equation}
where
$\mu(E, L) = \sigma_E\sigma_L |\Sigma|^{-1}g\left[E,L|\phi(r)\right]$,
and for the pdf we plug in Eq.~\eqref{eq:f_est} with ${d=2}$ and
${D_{ik} = \sqrt{(E_i' - E_k')^2 + (L_i' - L_k')^2}}$. If desired, we
replace
$g\left[E,L|\phi(r)\right] = (\sigma_L^2/\sigma_E) g\left[E', L'\Big
  |\phi(r')/(\sigma_L \sqrt{\sigma_E})\right]$, where
${r' = (\sqrt{\sigma_E}/\sigma_L)r}$.

\subsection{Generic integrable potential, $f = f(\vec{J})$}
\label{sec:f_J}
For realistic galactic potentials, assuming that most orbits are
regular or weakly chaotic, we may compute approximate actions with
e.g. the St\"ackel approximation \citep[][]{Binney2012}. In this
context, a system in dynamical equilibrium is described by a pdf in
action space
\begin{equation}
    F(\vec{J}) = (2\pi)^3f(\vec{J}),
    \label{eq:F_J}
\end{equation}
where $\vec{J}$ are three actions. Thus, Eq.~\eqref{eq:S_def_norm} reduces to
\begin{equation}
    S_{\vec{J}} = -\int F(\vec{J}) \ln \left[\frac{|\Sigma|F(\vec{J})}{(2\pi)^3} \right]\, \dd\vec{J}.
    \label{eq:S_J_def}
\end{equation}
The simplicity of Eq.~\eqref{eq:S_J_def}, in comparison to
Eqs.~\eqref{eq:S_E_def}-\eqref{eq:g_E} or
Eqs. \eqref{eq:S_EL_def}-\eqref{eq:T_r}, illustrates the advantages of
using action-based DFs instead of other integrals of motion. Defining
new actions $\vec{J}'$ normalized by their dispersions
$(\sigma_{J_1}, \sigma_{J_2}, \sigma_{J_3})$, we have
\begin{equation}
    \hat{S}_{\vec{J}} = -\frac{1}{N}\sum_{i=1}^N \ln \left[\frac{\hat{F_i}'(\vec{J}_i')}{\mu}\right],
    \label{eq:S_J_est}
\end{equation}
where
$\mu = (2\pi)^3\sigma_{J_1}\sigma_{J_2}\sigma_{J_3}|\Sigma|^{-1}$, and
for the pdf we plugin Eq.~\eqref{eq:f_est} with $d=3$ and
${D_{ik} = \sqrt{|\vec{J}_i' - \vec{J}_k'|^2}}$.

The same expressions apply to the cross-entropy estimates,
Eqs.~\eqref{eq:H_hat_norm}-\eqref{eq:xi_est}, \emph{mutatis mutandis}.

Having presented the expressions in general and for DFs depending only
on integrals of motion, in the next section we illustrate the physical
basis of the method, as well as investigate the bias and fluctuation
in these estimates. For that, we use a model with explicit expressions
for $f(E)$, $g(E)$ and for the actions.

\section{The isochrone model}
\label{sec:isochrone}
To illustrate the accuracy of these entropy estimators and the
physical basis of our method, we consider the isochrone model
\citep[][]{Henon1959}, whose potential is
\begin{equation}
    \phi(r) = -\frac{GM}{b}\frac{1}{1 + \sqrt{1 + (r/b)^2}},
\end{equation}
where $M$ is the total mass and $b$ the scale length. The DF of a
self-consistent sample is \citep[see][]{BT, Binney1985}
\begin{multline}
    f(E) = \frac{1}{\sqrt{2}(2\pi)^3(GMb)^{3/2}}\frac{\sqrt{\varepsilon}}{[2(1-\varepsilon)]^4}\times \Bigg[ 27 - 66\varepsilon + \\ 320\varepsilon^2 - 240\varepsilon^3 + 64\varepsilon^4 + 3(16\varepsilon^2 + 28\varepsilon - 9)\frac{\sin^{-1}\sqrt{\varepsilon}}{\sqrt{\varepsilon(1-\varepsilon)}} \Bigg],
    \label{eq:f_isochrone}
\end{multline}
and the density of states, Eq.~\eqref{eq:g_E}, is
\begin{equation}
    g(E) = (2\pi)^3\sqrt{GM}b^{5/2}\frac{(1-2\varepsilon)^2}{(2\varepsilon)^{5/2}},
    \label{eq:g_isochrone}
\end{equation}
where $\varepsilon=-bE/(GM)$. The radial period is
\begin{equation}
    T_r(E,L) = \frac{2\pi GM}{(-2E)^{3/2}}.
\end{equation}
As for any spherical system, the azimuthal and latitudinal actions are
${J_\varphi = L_z}$ and ${J_\theta = L - |L_z|}$, respectively, and
the radial action is
\begin{equation}
    J_r = \frac{1}{\pi}\int_{r_\mathrm{per}}^{r_\mathrm{apo}}\dd r\sqrt{2E - 2\phi(r) - L^2/r^2}.
\end{equation}
For the isochrone potential,
\begin{equation}
    J_r = \frac{GM}{\sqrt{-2E}} - \frac{1}{2}\Big( L + \sqrt{L^2 + 4GMb}\Big).
\end{equation}

\subsection{Entropy bias}
\label{sec:entropy_bias}

We start evaluating the integral in Eq.~\eqref{eq:S_E_def} numerically
with Eqs.~\eqref{eq:f_isochrone}-\eqref{eq:g_isochrone}, from
${E_\mathrm{min}=-0.5}$ to ${E_\mathrm{max}=-10^{-8}}$, with
$G = M = b = 1$. We take this as the true entropy value,
$S_\mathrm{E, true}$ -- thick solid grey line in
Fig.~\ref{fig:S_isochrone_self_cons} (upper panel). To compare with
the entropy estimates, we generate self-consistent samples with
different sizes $N$ of this model with \textsc{Agama}
\citep[][]{Vasiliev2019}, and integrate orbits for these samples for
$50\times\langle T_\mathrm{circ}\rangle$, where $T_\mathrm{circ}$ is
the period of circular motion. Fig.~\ref{fig:S_isochrone_self_cons}
(upper panel) shows the entropy estimates $\hat{S}_\mathrm{6D}$ (thin
solid lines) at different times and for different $N$ (colors), taking
the nearest neighbor ($k=1$). We recalculate
$|\Sigma|=\sigma_{w_1}\dots\sigma_{w_6}$, renormalizing the
coordinates at each time with the appropriate change of variables in
Eq.~\eqref{eq:S_def_norm}. This provides better estimates than a fixed
initial normalization, but the difference is small.

\begin{figure}
	\includegraphics[scale=0.425]{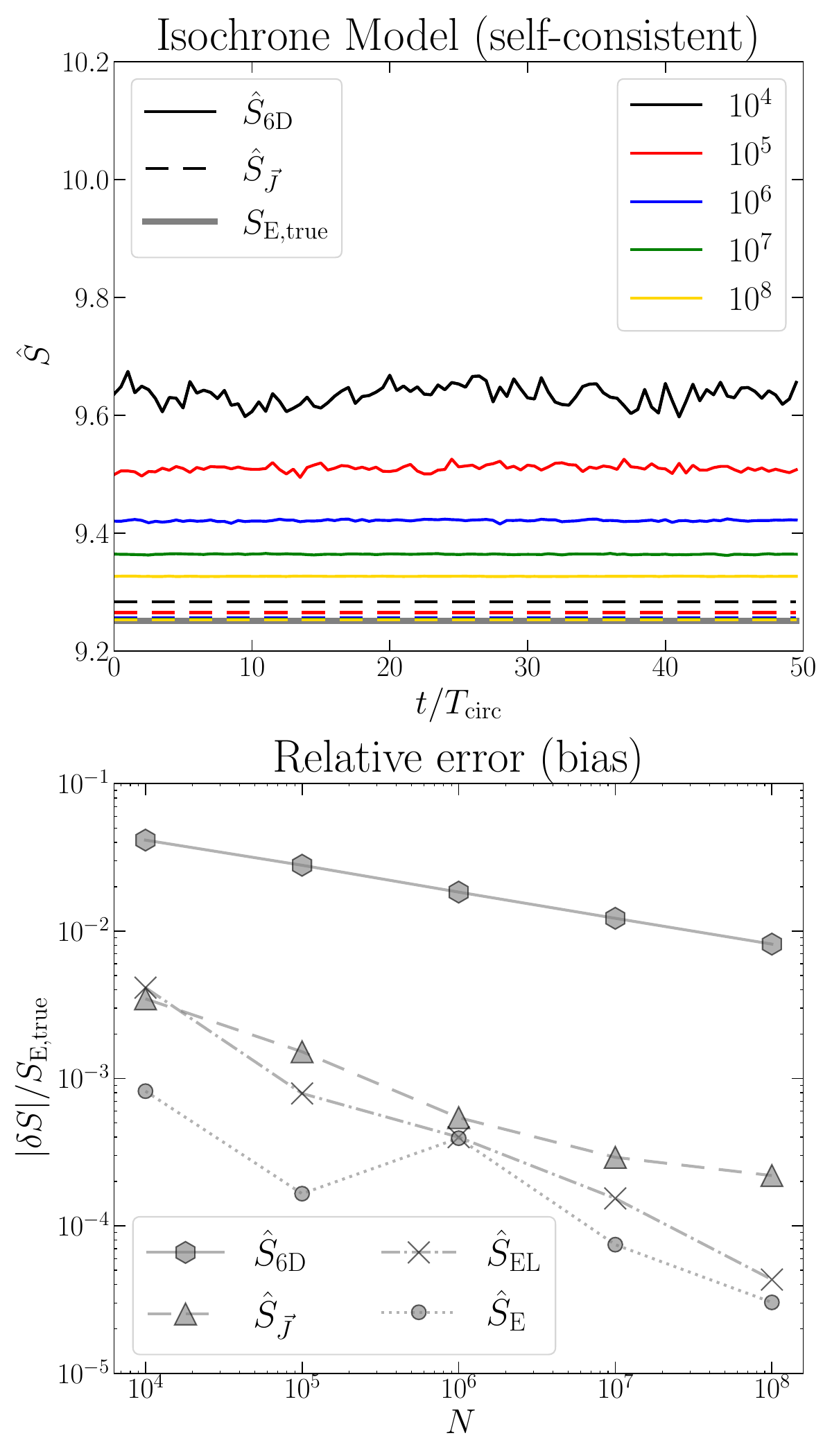}
        \caption{Top: entropy estimates in 6D (solid) and assuming the
          DF is an unknown function $f(\vec{J})$ (dashed) for
          self-consistent samples of the isochrone model, with
          different sample sizes (colors). The thick solid grey line
          shows the true value -- numerical integral in
          Eq.~\eqref{eq:S_E_def}. Bottom: relative error (bias) of
          $\hat{S}_\mathrm{6D}$, $\hat{S}_{\vec{J}}$,
          $\hat{S}_\mathrm{EL}$ and $\hat{S}_\mathrm{E}$. For a fixed
          sample size, estimates in lower dimensions are more
          accurate.}
    \label{fig:S_isochrone_self_cons}
\end{figure}

Since the initial sample is self-consistent with the potential, it is
stationary and $\hat{S}_\mathrm{6D}$ should be conserved. We see that
this is the case for all sample sizes, with larger fluctuations for
smaller $N$. Furthermore, $\hat{S}_\mathrm{6D}$ is significantly
biased with respect to the true value, and this bias is
time-independent, except for minor fluctuations. In the bottom panel,
the hexagons show the relative bias
$\delta S_\mathrm{6D} = (\langle \hat{S}_\mathrm{6D}\rangle_t -
S_\mathrm{E, true})/S_\mathrm{E, true}$ as a function of $N$, where
$\langle \hat{S}_\mathrm{6D}\rangle_t$ is a time-average. Even for
$N=10^8$, $\hat{S}_\mathrm{6D}$ has a relative bias of $\approx1\%$.

Fig.~\ref{fig:S_isochrone_self_cons} (upper panel) shows the entropy
estimates $\hat{S}_{\vec{J}}$, Eq.~\eqref{eq:S_J_est}, i.e. assuming
the DF is an unknown function of the actions (dashed lines). Since
these are conserved, we only estimate $S_{\vec{J}}$ at $t=0$. We see
that $\hat{S}_{\vec{J}}$ produces a much smaller bias, due to the
dimension reduction from 6D to 3D in the practical estimates -- but
$S_{\vec{J}}$ is still the entropy of the 6D DF. The triangles in the
bottom panel show that the bias stays below $\approx 1\%$ even for
$N=10^4$. Crosses and dots show the relative bias for
$\hat{S}_\mathrm{EL}$ and $\hat{S}_{E}$, respectively. These are
estimated with Eq.~\eqref{eq:S_E_est} for $S_\mathrm{E}$,
i.e. assuming the DF is an unknown function $f=f(E)$, and
Eq.~\eqref{eq:S_EL_est} for $S_{EL}$. We see that the bias is also
significantly smaller than that of $\hat{S}_\mathrm{6D}$, and it is
generally smaller for lower dimensions, as expected.

Thus, we have shown that: $\hat{S}_\mathrm{6D}$ is appropriately
conserved in the self-consistent model, but it is biased with respect
to the true value by $\delta S/S_\mathrm{E, true}\approx 5\%$ for
$N=10^4$, whereas in the space of integrals
$\delta S/S_\mathrm{E, true}< 1\%$ for $N=10^4$, and
$\delta S/S_\mathrm{E, true}\lesssim 0.01\%$ for $N=10^8$.

\subsection{Phase-mixing and entropy increase}
\label{sec:phase_mix_S_increase}
Here we use the same initial sample and consider two new isochrone
potentials with ${(M, b) = (3,1)}$ and $(M, b) = (1, 0.1)$, besides
the self-consistent one. Fig.~\ref{fig:hist_Jr} (top) shows histograms
of the radial actions $J_r$ evaluated in these three potentials. We do
not show histograms of $J_\varphi$ or $J_\theta$, since they do not
depend on the potential and are identical in the three cases. The
histogram is narrow in the original (self-consistent) potential and
broader in the new ones. The middle row shows histograms of the
angle-variables evaluated in the three potentials - all panels in this
figure have 50 equally spaced bins. Since the sample is not
phase-mixed in the new potentials, $\theta_r$ is not uniformly
distributed in these cases.

We then integrate orbits for this initial sample for
$50\times \langle T_\mathrm{circ}\rangle$ in the two new isochrone
potentials, which is long enough for the samples to
relax. Fig.~\ref{fig:hist_Jr} (bottom) shows histograms of the final
angles, as well as the initial ones in the self-consistent
potential. As expected for phase-mixed samples, these are all equally
uniform. In fact, Kolmogorov-Smirnov tests comparing the $\theta_r$
distribution in the self-consistent potential with the final ones in
the new potentials result in statistic values $\sim 0.01$, with
p-values $\gtrsim 0.6$, largely failing to reject the equally-uniform
hypothesis.
\begin{figure}
	\includegraphics[width=\columnwidth]{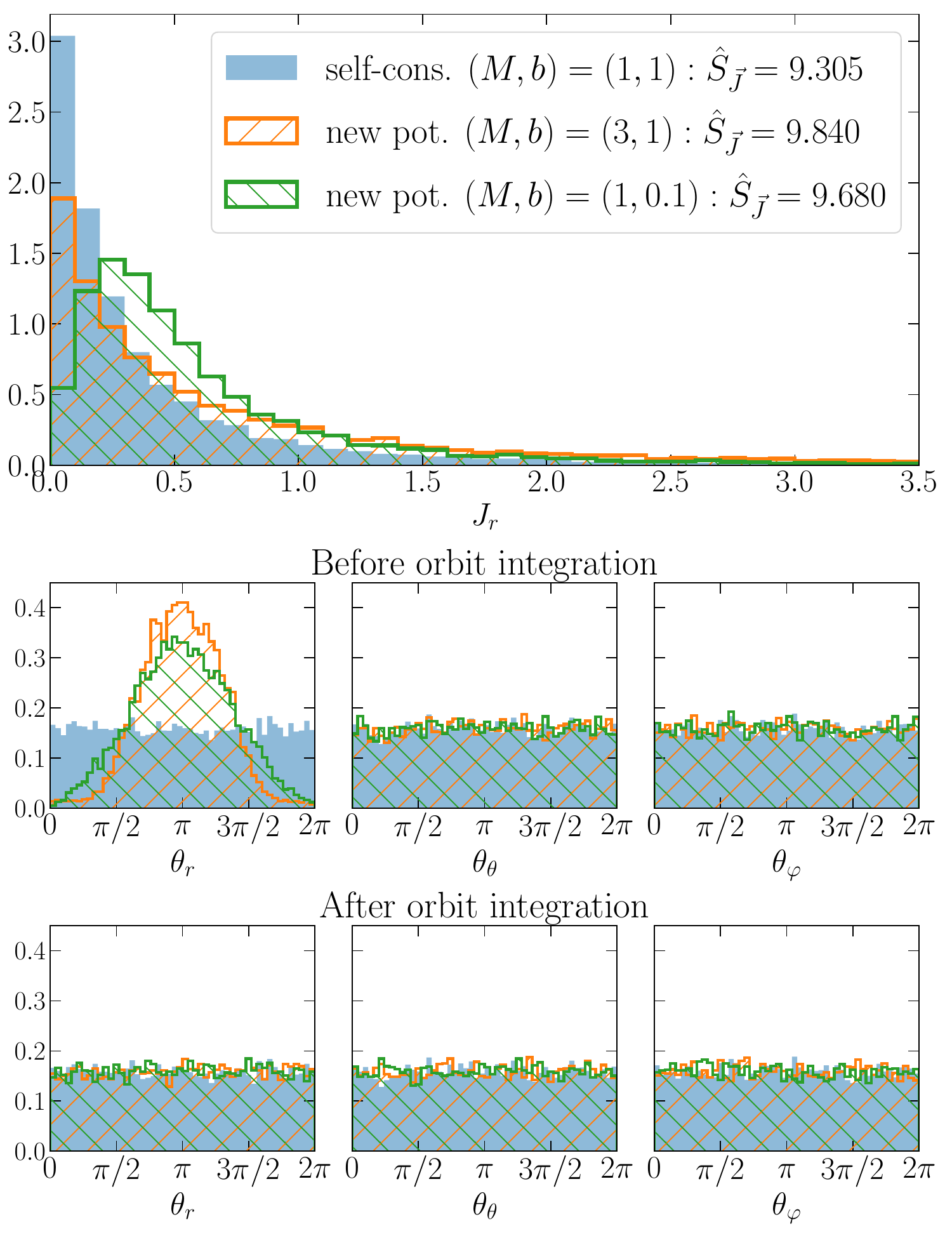}
        \caption{Top: histograms of the radial action $J_r$ for a
          self-consistent sample of an isochrone model with
          $(M,b)=(1,1)$, with $J_r$ evaluated in this model
          (``self-cons.'') and for $(M, b) = (3,1)$ and
          $(M,b) = (1, 0.1)$. The distribution is broader in the new
          potentials. Middle: histograms of angle-variables for the
          original sample in the self-consistent potential and in the
          other ones (where the original sample is not
          stationary). Bottom: same as the middle row, but after orbit
          integration in each of the new potentials. The final angle
          distributions are all uniform, as expected for phase-mixed
          samples. The legend shows final entropy values in each
          potential. All panels have 50 equally spaced bins.}
    \label{fig:hist_Jr}
\end{figure}
Just as for the self-consistent sample, this uniformity does not
require any coarse-graining, but is a rather objective fact.

We now show, before estimating the entropy, that the evolution of the
original sample in new potentials, as illustrated in
Fig.~\ref{fig:hist_Jr}, is necessarily accompanied by an entropy
increase. We define the sample's initial entropy as
$S[f_0] = -\int f_0 \ln f_0 \dd \vec{w} = - \int f_0 \ln f_0
\dd\vec{\theta}\dd\vec{J}$, where $f_0(\vec{w})$ is the initial DF and
$\vec{w} = (\vec{r}, \vec{v})$. $S[f_0]$ is invariant for
angle-actions evaluated in any potential. For the self-consistent
potential, $f_0(\vec{\theta}, \vec{J}) = (2\pi)^{-3}F_0(\vec{J})$ and
thus
\begin{equation}
    S_{\vec{J}}[f_0] = \ln (2\pi)^3 + S[F_0], 
    \label{eq:S_f_0}
\end{equation}
where $S[F_0] = -\int F_0(\vec{J}) \ln F_0 \dd\vec{J}$. Similarly,
after phase-mixing in the new potential, the final DF is
$f_\mathrm{final}(\vec{\theta}, \vec{J})=
(2\pi)^{-3}F_\mathrm{final}(\vec{J})$ and its entropy is
\begin{equation}
    S_{\vec{J}}[f_\mathrm{final}] = \ln (2\pi)^3 + S[F_\mathrm{final}]. 
    \label{eq:S_f_final}
\end{equation}
Since the three samples have the same actions' distribution, except
for $J_r$ being broader in the new potentials
(Fig.~\ref{fig:hist_Jr}), we see that
$S[F_\mathrm{final}]>S[F_\mathrm{0}]$, for broader pdfs have larger
entropies. Thus, from Eqs.~\eqref{eq:S_f_0} and \eqref{eq:S_f_final},
$S_{\vec{J}}[f_\mathrm{final}]>S_{\vec{J}}[f_\mathrm{0}]$, i.e. the
phase-mixing of a non-relaxed sample is necessarily accompanied by an
entropy increase. This is confirmed by our estimates (legend). We
emphasize that the practical entropy calculation only uses actions,
while assuming that the final angle distribution will be uniform, as
required by Jeans' theorem. For a given sample in any trial potential,
we can estimate the final entropy right away, since actions are
conserved.

To study the sample evolution in more detail,
Fig.~\ref{fig:S_isochrone_3M1} (upper panel) shows entropy estimates
using 6D coordinates at several time-steps ($\hat{S}_\mathrm{6D}$,
solid lines) as well as $\hat{S}_{\vec{J}}$ (dashed) for the same
initial sample evolved in the potential $(M, b) = (3,1)$. Since the
initial sample is not in dynamical equilibrium in the new potential,
it responds to the higher mass developing a radially biased velocity
anisotropy. The final DF is unknown, but it should respect Jeans'
theorem, being a function $f(E,L)$, or $f(\vec{J})$. The thick solid
grey line shows $\hat{S}_\mathrm{EL}$ for $N=10^8$ in the new
potential, which is the lower dimension allowed by the phase-mixed
sample. Here we proceed exactly as previously to get
$\hat{S}_\mathrm{EL}$ for the self-consistent sample
(Fig.~\ref{fig:S_isochrone_self_cons}), the only difference being that
energies are evaluated in a new potential. Since we have shown that
$\hat{S}_\mathrm{EL}$ has a negligible bias for $N=10^8$, we take this
as the true final entropy, $\hat{S}_\mathrm{EL, true}$.

\begin{figure}
	\includegraphics[scale=0.425]{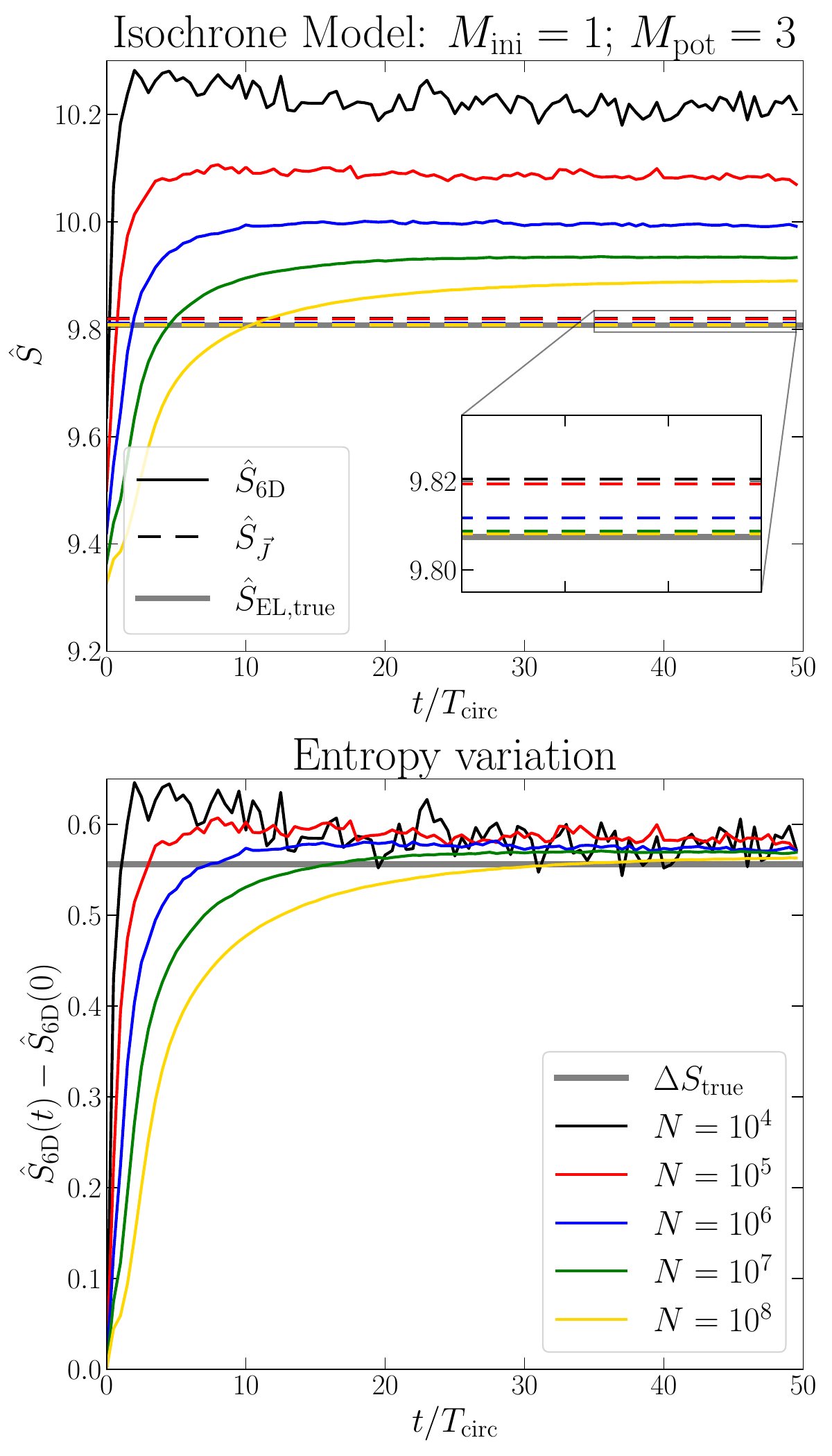}
        \caption{Top: entropy estimates in 6D (solid) and assuming
          $f=f(\vec{J})$ (dashed) for initial self-consistent samples
          of the isochrone model with $M=1$, but integrated in (and
          $\vec{J}$ evaluated at) an isochrone potential with
          $M=3$. The thick solid grey line shows the entropy for a
          phase-mixed system with $f=f(E,L)$ and $N=10^8$, considered
          as the true final entropy. Bottom: entropy variation
          $\Delta \hat{S} = \hat{S}_\mathrm{6D}(t) -
          \hat{S}_\mathrm{6D}(0)$ for different sample sizes, which
          approximately converges to
          $\Delta S_\mathrm{true} = \hat{S}_{\mathrm{EL, true}} -
          S_{\mathrm{E, true}}$ for all samples.}
    \label{fig:S_isochrone_3M1}
\end{figure}

Besides the biases with respect to the initial true entropy
$S_\mathrm{E, true}$ (Fig.~\ref{fig:S_isochrone_self_cons}),
Fig.~\ref{fig:S_isochrone_3M1} shows that the asymptotic values of
$\hat{S}_\mathrm{6D}$ ($t\rightarrow\infty$) in the new potential are
also biased with respect to $\hat{S}_\mathrm{EL, true}$. On the other
hand, $\hat{S}_{\vec{J}}$ is again much less biased, since it is
estimated in a lower dimension space, while the angles' contribution
to $\hat{S}_{\vec{J}}$ is $\ln(2\pi)^3$ -- see
Eq.~\eqref{eq:S_f_final}. In both cases the bias decreases for larger
$N$ -- see the inset plot.

Fig.~\ref{fig:S_isochrone_3M1} (bottom) shows
${\Delta \hat{S}_\mathrm{6D} = \hat{S}_\mathrm{6D}(t) -
  \hat{S}_\mathrm{6D}(0)}$ (colors) and
${\Delta S_\mathrm{true} = \hat{S}_\mathrm{EL, true} - S_\mathrm{E,
    true}}$, the true entropy increase (thick grey). The final
$\Delta \hat{S}_\mathrm{6D}$ is similar for all sample sizes,
approximately converging to $\Delta S_\mathrm{true}$. This confirms
that the bias is nearly independent of time and is thus nearly
eliminated by calculating entropy variations, as done by
\cite{BeS2017,BeS2019a,BeS2019b}.

\subsection{Bias correction}
\label{sec:bias_correction}
If the bias of $\hat{S}$ does not depend on the model parameters, it
poses no problem for the minimum-entropy fits, since it only
introduces an additive constant in $\hat{S}$. For a possibly
model-dependent bias, we investigate it in more detail and test a
prescription to suppress it.

It is known that taking the $k$-th neighbor for larger $k$ increases
the bias in the entropy estimate, but decreases its variance, a
manifestation of the bias-variance trade-off
\citep[e.g.][]{Wasserman2010}. To investigate this, we generate $10^3$
realizations of size-$N$ self-consistent isochrone samples with
${M=b=1}$ and calculate actions and $\hat{S}_{\vec{J}}$,
Eq.~\eqref{eq:S_J_est}, for each realization, normalizing the actions
in each one. Here we do not compare with $\hat{S}_\mathrm{6D}$, thus
we do not normalize by $|\Sigma|=\sigma_{w_1}\dots\sigma_{w_6}$, which
would introduce unnecessary extra noise.

Fig.~\ref{fig:variance} (upper panel) shows the bias, i.e. the
difference between the mean of the realizations and the true value, as
a function of $k$ for different sample sizes (full triangles). We
confirm the increase in the bias for larger $k$, with $k=10$ producing
a $\sim 2\times$ larger bias than $k=1$.

\begin{figure}
	\includegraphics[width=\columnwidth]{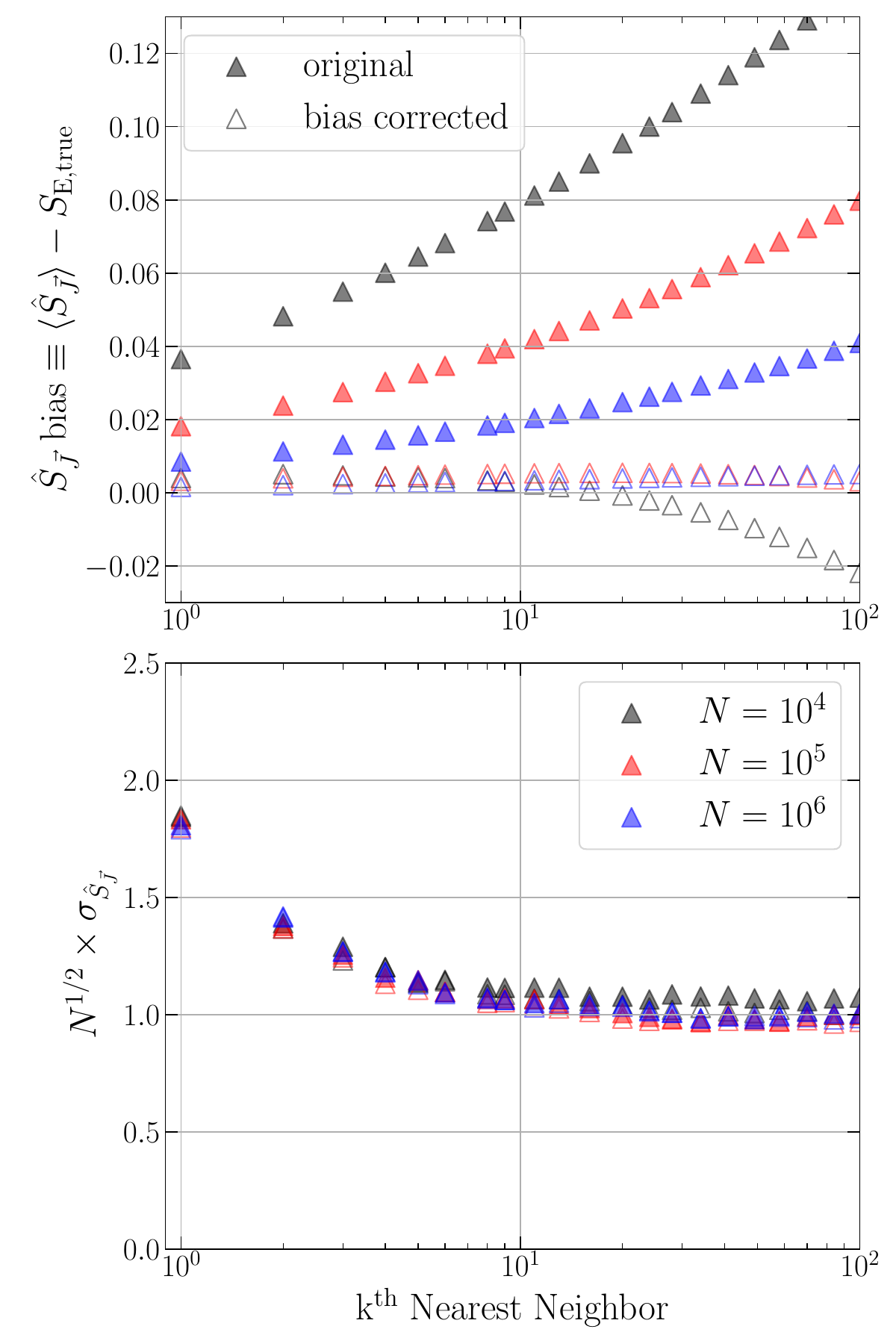}
        \caption{Bias (top) and fluctuation (bottom) of entropy
          estimates for self-consistent samples of the isochrone
          model. We see that the uncorrected bias (full triangles)
          increases with $k$, with the correction suppressing the
          bias. Empty blue and red triangles nearly overlap. The
          fluctuation $\sigma_{\hat{S}_{\vec{J}}}$ decreases with $k$,
          saturating at
          $\sigma_{\hat{S}_{\vec{J}}} \approx 1/\sqrt{N}$ for
          $k\approx 10$.}
    \label{fig:variance}
\end{figure}

We investigate the correction of \cite{Charzynska2015}, who suggest
that the bias is essentially due to points near the edges of the
distribution support. For these points, the hyper-sphere around the
point (defined by the distance $D_{ik}$ to the $k$-th neighbor) can
have a fraction of its volume outside the support. This results in
overestimating the volume, and Eq.~\eqref{eq:f_est} underestimating
the DF for these points. When plugged into Eq.~\eqref{eq:S_hat_norm},
this produces a positive bias, in accordance with our results --
Figs.~\ref{fig:S_isochrone_self_cons}, \ref{fig:S_isochrone_3M1} and
\ref{fig:variance}. To compensate for this, \cite{Charzynska2015}
propose to add the following correction to the entropy estimate:
\begin{equation}
    C = \frac{1}{N} \sum_{i=1}^{N} \ln \left( \frac{|v(\vec{w}_i, D_{ik}) \cap \mathrm{supp(W)}|}{|v(\vec{w}_i, D_{ik})|}\right),
    \label{eq:bias_correc}
\end{equation}
where $v(\vec{w}_i, D_{ik})$ is the volume around point $\vec{w}_i$,
which is drawn from $W$, in $d$-dimensions.

In general, the support's shape and the intersections in
Eq.~\eqref{eq:bias_correc} are unknown, and \cite{Charzynska2015}
propose assuming a hyper-rectangular box for the support and a
hyper-cubic box for the volume $v(\vec{w}_i, D_{ik})$, although their
analysis restricted to $k=1$. For cubic boxes of side $l_i$, we
correct for points such that $w_{j,i} > w_{j,\mathrm{max}} - l_i/2$,
or $w_{j,i} < w_{j,\mathrm{min}} + l_i/2$, where $j=1,\dots,d$, and
calculate the volume fractions of the cube inside the rectangular
box. Concisely, it results
\begin{equation}
\begin{aligned}
    C = \frac{1}{N} \sum_{i=1}^{N} \sum_{j=1}^{d}\ln \Bigg[ & \mathrm{min}\Big(\frac{w_{j,\mathrm{max}}}{l_i}, \frac{w_{j,i}}{l_i} + \frac{1}{2}\Big) - \\
    & \mathrm{max}\Big(\frac{w_{j,\mathrm{min}}}{l_i}, \frac{w_{j,i}}{l_i} - \frac{1}{2}\Big)\Bigg].
\end{aligned}
    \label{eq:bias_correc_cubic}
\end{equation}
After a few experiments, we settled a cube inscribed within the sphere
of radius $D_{ik}$, i.e. $l_i =
(2/\sqrt{d})D_{ik}$. Fig.~\ref{fig:variance} (top) shows the corrected
biases (empty triangles), which are smaller than the original ones by
factors $5-15$ -- note that the empty blue and red triangles nearly
overlap. The improvement is even better for larger $k$, where the bias
is not larger than that of $k=1$ (up to some $k$, beyond which the
bias is over-corrected).

\subsection{Entropy fluctuation}
\label{sec:entropy_fluctuation}
The noise in the entropy estimate is theoretically expected to have
amplitude ${\sigma_{\hat{S}} \approx N^{-1/2}}$
\citep[][]{Biau2015}. Fig.~\ref{fig:variance} (bottom) shows the
fluctuations ${\sigma_{\hat{S}_{\vec{J}}}}$, estimated as half the
$16-84$th-interpercentile range of the realizations, and multiplied by
$N^{1/2}$. We confirm that
${\sigma_{\hat{S}_{\vec{J}}} \approx N^{-1/2}}$, and we see that
$\sigma_{\hat{S}}$ decreases with $k$, but it saturates at
$k\approx 10$, reducing $\sigma_{\hat{S}_{\vec{J}}}$ by a factor
$\approx 2$ in comparison to $k=1$. Empty triangles show
${\sigma_{\hat{S}_{\vec{J}}}}$ for the bias-corrected estimates, which
are nearly identical to those of the uncorrected estimates.

In summary, we conclude that taking $k=10$ suppresses the noise by a
factor 2, and the correction proposed by \cite{Charzynska2015}
suppresses the bias without increasing the noise.

\section{Minimum entropy illustrated}
\label{sec:min_entropy_illustra}
In Appendix \ref{sec:appendix}, we rigorously demonstrate why the
entropy of a fixed sample is minimum in the correct potential, i.e. in
the one where the sample is phase-mixed. In this section, we
illustrate this with phase-mixed samples in a self-consistent
isochrone model and in potentials of the hypervirial family
\citep[][]{Evans2005}.

\subsection{Isochrone potential}
We generate an initial sample of the isochrone model with $M=b=1$, and
sample size $N/0.7$, selecting the $70\%$ most bound particles in the
self-consistent potential, with a final sample of $N\approx
10^4$. This allows us to explore a larger set of models, since we
restrict to models where all particles are bound. Note that this cut
does not affect the method because the DF is still a function of
integrals of motion only, and self-consistency is not required as we
explicitly demonstrate below.

\begin{figure}
\begin{center}
	\includegraphics[scale=0.4]{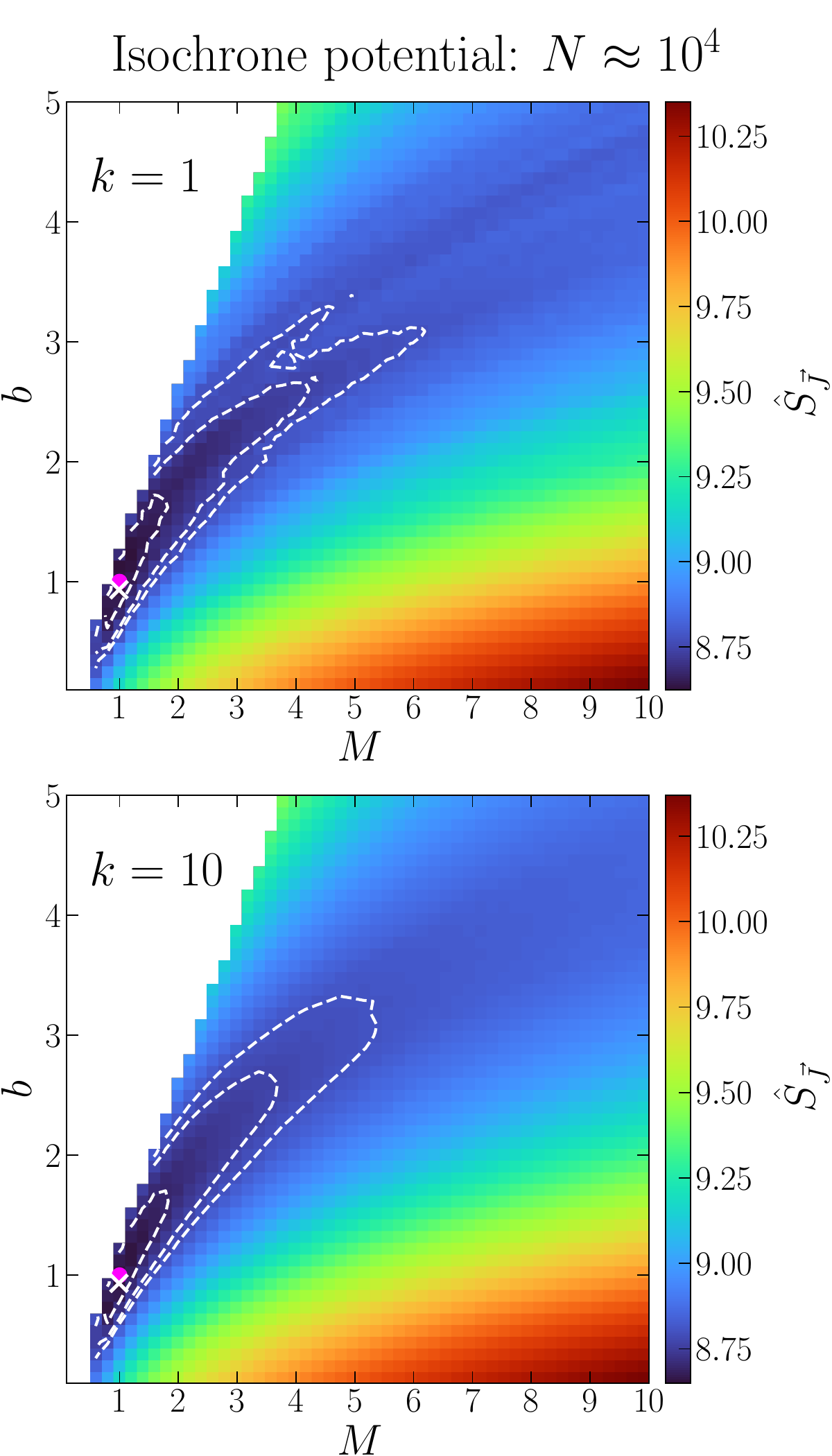}
        \caption{$\hat{S}_{\vec{J}}$ values of a self-consistent
          sample of the isochrone model $(M, b)=(1,1)$ (magenta dots)
          with actions evaluated on a grid of parameters ($M,b$) of
          the isochrone potential. Contours are percentile levels
          relative to the minima of $\hat{S}_{\vec{J}}$ (white
          X's). $\hat{S}_{\vec{J}}$ is estimated with the
          nearest-neighbor, $k=1$ (top), and $k=10$ (bottom). As
          expected, a larger $k$ smooths out the
          $\hat{S}_{\vec{J}}$-surface
          (Sec.~\ref{sec:entropy_fluctuation}). $\hat{S}_{\vec{J}}$ is
          minimum near the true potential where the sample is
          phase-mixed.}
    \label{fig:S_grid_isoc}
\end{center}
\end{figure}

We calculate $\hat{S}_{\vec{J}}$, i.e. the entropy the sample would
have after phase-mixing, on a grid of potentials $(M,b)$, but in this
exercise we do not correct the bias discussed in
Sec.~\ref{sec:bias_correction}. Fig.~\ref{fig:S_grid_isoc} shows
$\hat{S}_{\vec{J}}$ values in the grid ($M, b$), using the nearest
neighbor, $k=1$ (top), and $k=10$ (bottom). The magenta dots show the
true parameters, and the white X's show the location of the minimum
entropy. The minimum entropy is indeed very near the true values. We
note, however, that its exact location depends on the sample
realization. The white curves are illustrative contours of the 1st,
5-th and 10-th percentiles of $\hat{S}_{\vec{J}}$ (not credible
contours). The wrinkles in the colors and contours in the top panel
reveal the noise in $\hat{S}_{\vec{J}}$ for $k=1$, while for $k=10$
the surface is much smoother, in agreement with
Fig.~\ref{fig:variance} (bottom panel).

\subsection{Hypervirial potentials}
At this point, the reader might think that the identification of the
potential with the minimum entropy depends on something special about
the isochrone potential, or on having a self-consistent sample, as
opposed to a generic stationary sample. To dispel this concern, we now
use the same sample used before as initial conditions and integrate
orbits in four different potentials of the hypervirial family
\citep[][]{Evans2005} characterized by the potential/density pair
\begin{equation}
    \phi(r) = -\frac{GM}{a}\frac{1}{\left[1+\left(r/a\right)^p\right]^{1/p}},
\end{equation}
\begin{equation}
    \rho(r) = \frac{(p+1)M}{4\pi a^3}\frac{\left(r/a\right)^{p-2}}{\left[1+\left(r/a\right)^p\right]^{2+1/p}},
\end{equation}
where $0 < p \leq 2$ for the most physically interesting cases. These
models have $\rho \sim r^{p-2}$ near the center and
$\rho \sim r^{-(p+3)}$ in the outskirts, and have finite mass
$M$. Their most interesting property is that they respect the virial
theorem locally, besides the usual global one. We use these models for
their simplicity and because they reduce to well known models for
$p=1$ \citep{Hernquist1990}, and $p=2$ \citep{Plummer1911}. We also
explore the cases $p=1/2$ (strong cusp) and $p=3/2$ (weak cusp). We
set $G = a =1$, but $M=2$ in order to have only bound orbits in all
models. We integrate orbits for
$100\times \langle T_\mathrm{circ}\rangle$, which is enough for the
samples to phase-mix in each of these four potentials. This creates,
for each potential, a different equilibrium (phase-mixed) DF, with no
explicit expression. Then, for each of these four phase-mixed samples,
we calculate actions and $\hat{S}_{\vec{J}}$ in trial potentials
$(M,a)$, with the corresponding parameter $p$ fixed.

\begin{figure*}
\begin{center}
	\includegraphics[scale=0.4]{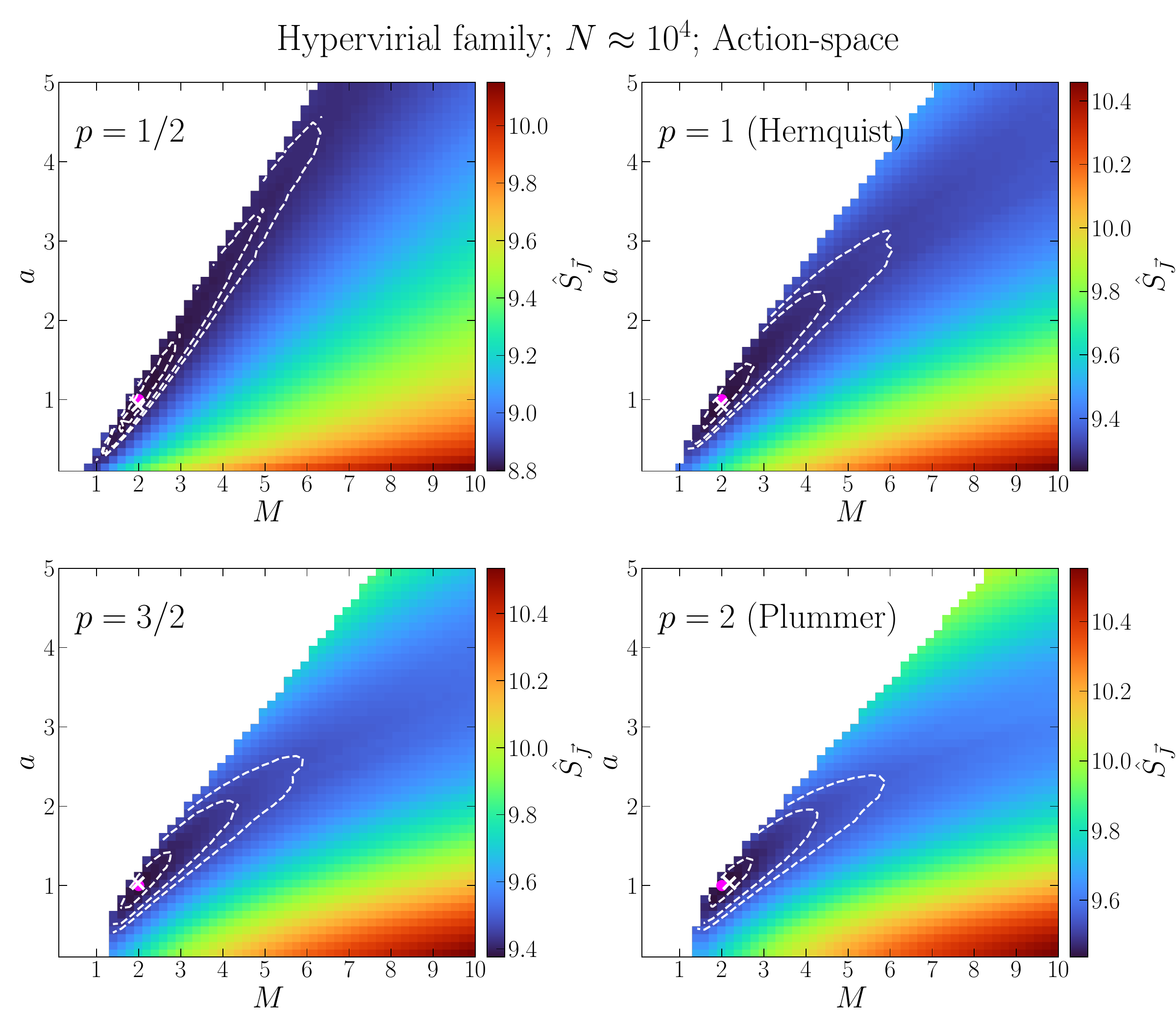}
        \caption{The $\hat{S}_{\vec{J}}$-surface with actions
          calculated on a grid of parameters ($M,a$). Each panel is
          for a different potential of the hypervirial family of
          \cite{Evans2005}, where the same initial sample
          phase-mixed. Magenta dots show the true values, with
          contours showing percentile levels relative to the minimum
          of $\hat{S}_{\vec{J}}$ (white X). The entropy is estimated
          with the $k=10$ nearest-neighbor. For all models,
          $\hat{S}_{\vec{J}}$ has its minimum near the correct
          parameters.}
    \label{fig:S_grid_hypervirial_J}
\end{center}
\end{figure*}

Fig.~\ref{fig:S_grid_hypervirial_J} shows the entropy for these
potentials. The minima (white crosses) lie near the true values
(magenta dots), but once more their exact locations depend on the
particular data realization. This shows that the only requirement of
the minimum-entropy method is that the sample is phase-mixed in the
true potential, with self-consistency playing no special role. Let us
emphasize that this procedure does not require knowing the sample's
density or anisotropy profile, or its DF, but only assumes that the DF
is an unknown function satisfying the Jeans' theorem,
i.e. $f=f(\vec{J})$.

Fig.~\ref{fig:S_grid_hypervirial_EL} shows a similar picture, but with
the entropy calculated in the space of energy and angular momentum,
Eqs.~\eqref{eq:g_EL}-\eqref{eq:S_EL_est}, with $T_r=2\pi/\Omega_r$,
where $\Omega_r$ is the radial frequency calculated with
\textsc{Agama}. Once more the entropy minima are close to the true
values for all models. This $\hat{S}_\mathrm{EL}$ is slightly noisier
than $\hat{S}_{\vec{J}}$, even though the former is defined in 2D and
thus expected to have smaller noise. We suspect that this extra noise
in $\hat{S}_\mathrm{EL}$ may be due to the numerical calculation of
the radial period in the density of states,
Eqs.~\eqref{eq:g_EL}-\eqref{eq:T_r}, further illustrating the
advantages of actions.

\begin{figure*}
\begin{center}
	\includegraphics[scale=0.4]{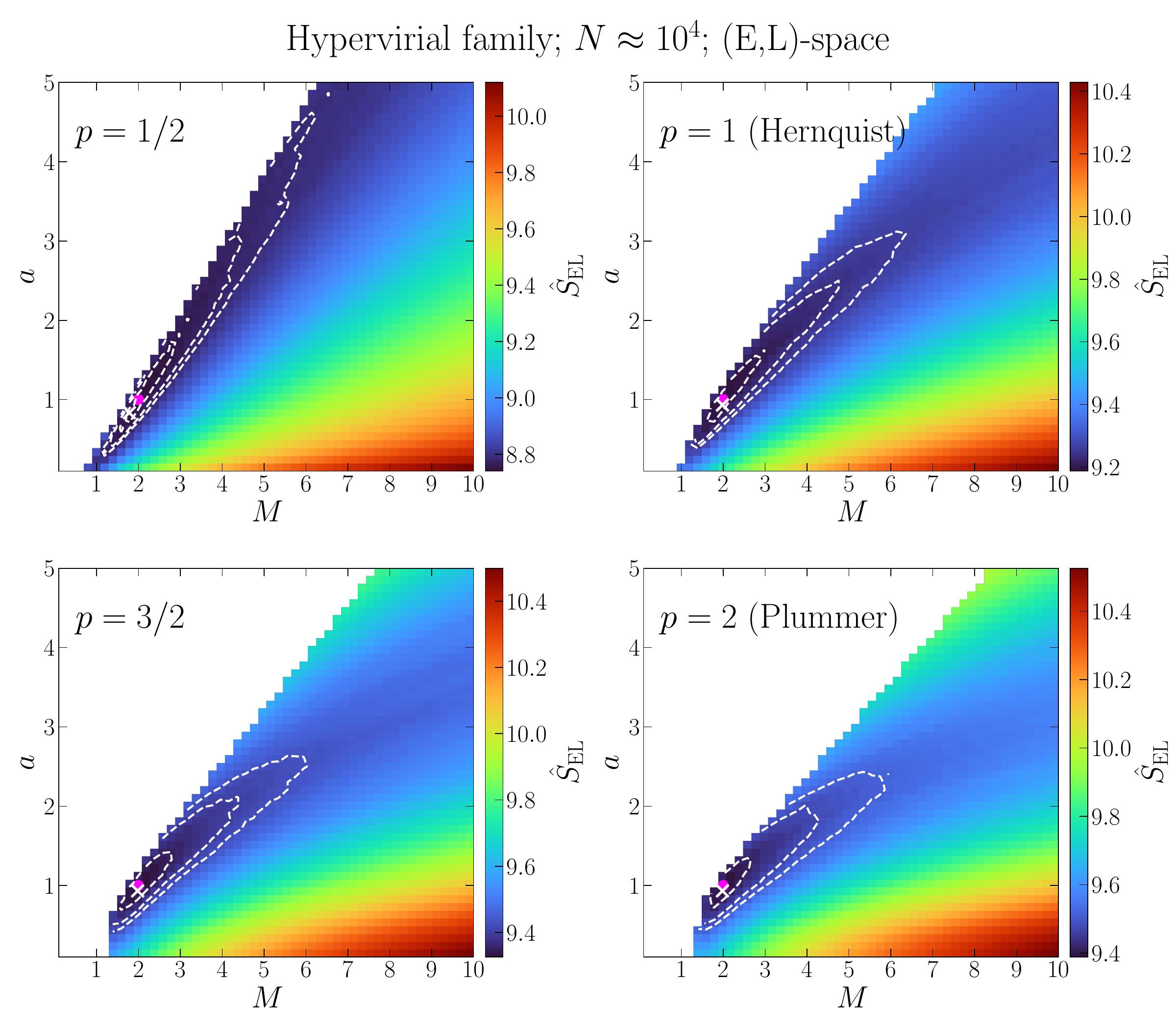}
        \caption{Similar to Fig.~\ref{fig:S_grid_hypervirial_J}, but
          calculated in the (E,L)-space. The minima of $\hat{S}_{E,L}$
          (white X) again are close to the true values (magenta dots),
          but the $\hat{S}_{E,L}$-surface has more wrinkles, revealing
          a sligthly larger noise.}
    \label{fig:S_grid_hypervirial_EL}
\end{center}
\end{figure*}

\section{Model fitting}
\label{sec:fits}

While Figs.~\ref{fig:S_grid_isoc}-\ref{fig:S_grid_hypervirial_EL} may
be seen as approximate fits, evaluating models in a grid can quickly
become inefficient for models with larger numbers of
parameters. Moreover,
Figs. ~\ref{fig:S_grid_isoc}-\ref{fig:S_grid_hypervirial_EL} do not
provide the odds ratios of different trial potentials. In this
section, we do actual fits in two steps. We first use the downhill
simplex (``Nelder-Mead''), as implemented in scipy, to minimize the
entropy of the final DFs considered as unknown functions $f(\vec{J})$,
with actions evaluated in trial potentials.

Having found the best-fit potential where the final (equilibrium) DF
is an unknown function $f_0(\vec{J})$, we explore the posterior of the
parameters of the potential. For that, one might want to use the
Kullback-Leibler divergence (KLD) between $f_0(\vec{J})$ and a trial
potential with final DF $f(\vec{J})$. We do not use the KLD as a
direct estimate of posterior ratios, as done by \cite{Sanderson2015},
but we present the main expressions in that approach for
completeness. The KLD is defined as
\begin{equation}
    D_\mathrm{KL}(f_0 || f) \equiv \int f_0 \ln \left( \frac{f_0}{f}\right) \, \dd\vec{\theta}\dd\vec{J} = H(f_0, f) - S_0,
    \label{eq:KLD}
\end{equation}
where $S_0 = S_{\vec{J}}[f_0]$. For two distributions $f$ and $g$ in
general, $D_\mathrm{KL}(f || g)$ can be seen as a directed distance
from $f$ to $g$. In fact, $D_\mathrm{KL}(f || f) = 0$, and it can be
shown that $D_\mathrm{KL}(f || g) \geq 0$ \citep{Kullback1968}. As for
the entropy, the KLD can be estimated via Monte-Carlo using samples of
$f$ and $f_0$, with no explicit expressions for these DFs.

From Eq.~\eqref{eq:lnL_expect}, we get \citep[e.g.][]{Cover_Thomas}
\begin{equation}
    D_\mathrm{KL}(f_0 || f) = \frac{1}{N}\left( \ln \Lcal_0  - \ln \Lcal\right),
    \label{eq:KLD_2}
\end{equation}
where $\ln \Lcal_0 = -N S_0$ is the expectation value of the
log-likelihood of the best model. From Bayes' theorem:
\begin{equation}
P(\vec{p}|\vec{w}) = \frac{\Lcal(\vec{w}|\vec{p})P(\vec{p})}{P(\vec{w})},
 \label{eq:Bayes}
\end{equation}
where $P(\vec{p}|\vec{w})$ is the parameters' posterior probability,
$P(\vec{p})$ is their prior probability and $P(\vec{w})$ is a
normalization factor. For the best-fit model $\vec{p}_0$:
\begin{equation}
P(\vec{p}_0|\vec{w}) = \frac{\Lcal(\vec{w}|\vec{p}_0)P(\vec{p}_0)}{P(\vec{w})}.
 \label{eq:Bayes_0}
\end{equation}
Taking the logarithm of Eqs.\eqref{eq:Bayes}-\eqref{eq:Bayes_0} and
replacing in Eq.~\eqref{eq:KLD_2}, one can translate the KLD into
ratios of posterior probabilities of different models. In particular,
for flat priors we have $P(\vec{p}) = P(\vec{p}_0)$.

However, this would make a point-wise comparison using a fixed sample
evaluated in different models, which does not take into account the
intrinsic uncertainties in the data-generation process. In other
words, to get meaningful posteriors one needs to recognize that the
data set is just a particular realization of underlying unknown DFs
$f_0$ and $f$. Not considering this and using KLD with a single sample
would produce unrealistically tiny credible contours. On the other
hand, neglecting the factor $1/N$ in Eq.~\eqref{eq:KLD_2}, as done by
\cite{Sanderson2015}, significantly overestimates the uncertainties.

We emphasize that we do not use the KLD as a direct translation of the
posterior probability ratios, but we use it as a distance metric to
explore the posterior probabilities in an Approximate Bayesian
Computation, as described below. In this way, each model is
accompanied by a different data realization and uncertainties in the
data-generation process are appropriately incorporated.

\subsection{Fitting the isochrone potential}
\label{sec:fit_isoc}
We generate a self-consistent sample of the isochrone model with
$M=b=1$, and sample size $N/0.7$, selecting the $70\%$ most bound
particles, with a final sample of $N\approx 10^4$. We assume that the
final DF describing the sample in each trial isochrone potential, if
orbits were integrated until phase-mixed, is an unknown function
$f(\vec{J})$. We estimate $S_{\vec{J}}$, Eq.~\eqref{eq:S_J_est},
taking the $k$-th neighbor with $k=10$ and correcting for the bias as
discussed in Sec.~\ref{sec:entropy_fluctuation}. We then minimize
$\hat{S}_{\vec{J}}$.

To prevent trapping at local minima, we fit the data starting with
initial parameters in a regular grid of $4\times 4$ points, with
$0.1 < M < 10$, and $0.1 < b <5$. We only fit potentials with no
unbound particle, setting ${\hat{S}_{\vec{J}} = \infty}$
otherwise. The best-fit potential, i.e. the one with smallest
$\hat{S}_{\vec{J}}$ among all fits, is shown as a red dot in
Fig.~\ref{fig:fit_isoc_100_boots}.
 
Having found the best-fit potential, we generate 100 new data sets via
bootstraps (randomly selecting $N$ points with replacement), fitting
the potential for each one. In principle, bootstrap samples might put
a problem in the entropy estimate, since duplicated points would have
zero-distance to the nearest neighbor. The solution, already
implemented in \texttt{tropygal}, is treating repeated points as
copies of the same point and neglecting copies in the search for
neighbors.

In Fig.~\ref{fig:fit_isoc_100_boots}, black points show results
obtained for the bootstrap samples, and the green triangle is the
median of the best-fit parameters. The parameters recovered with both
the original sample and with the median of the samples are very near
the true values, but we remark that they vary for different data
realizations.
 
\begin{figure}
\begin{center}
	\includegraphics[scale=0.4]{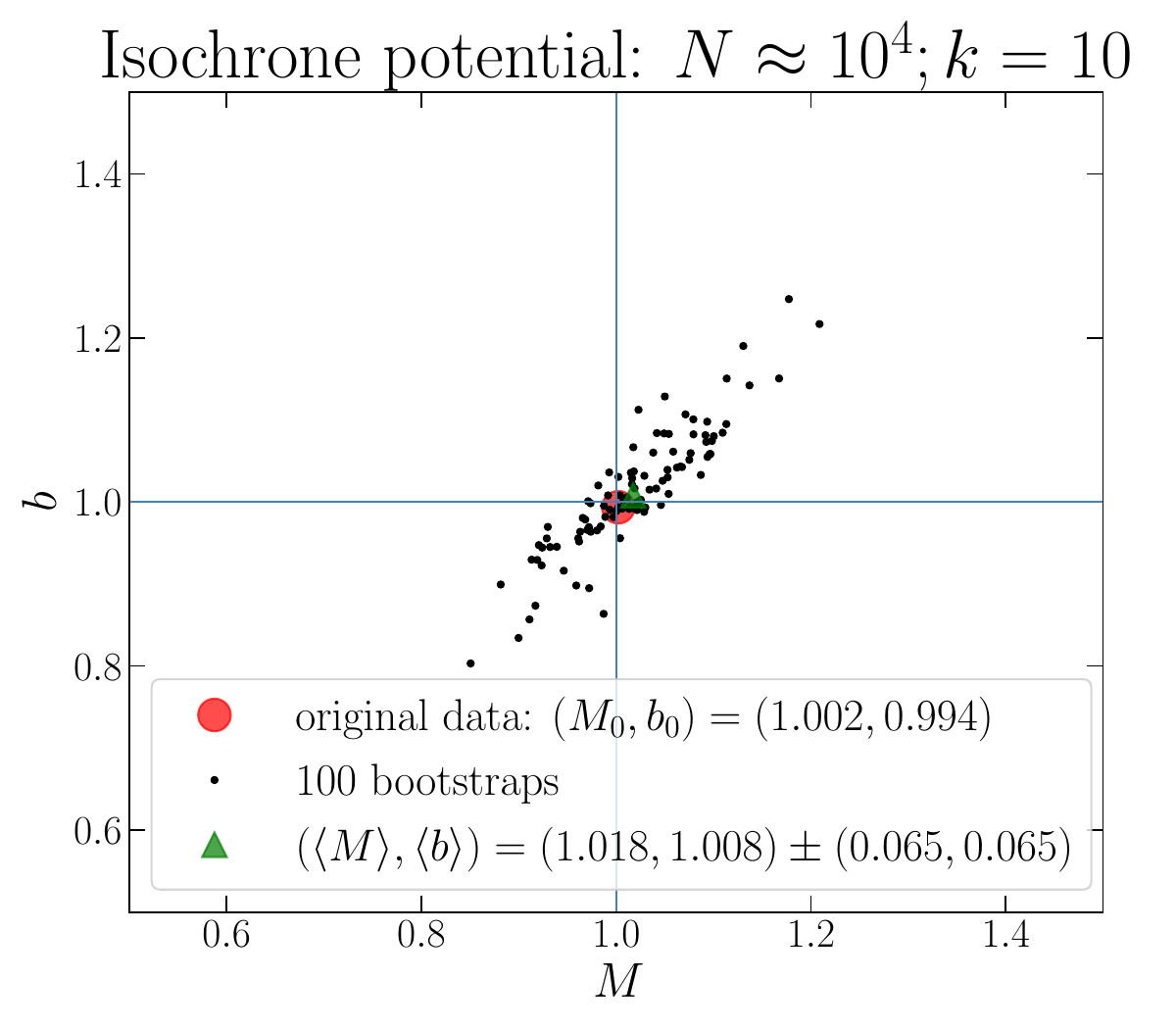}
        \caption{Fit results obtained with a sample of $N\approx 10^4$
          in equilibrium (red dot) and bootstrap samples (black
          points). The green triangle shows the median of the best-fit
          parameters. In both cases, the true values (lines) are well
          recovered.}
    \label{fig:fit_isoc_100_boots}
\end{center}
\end{figure}

In Fig.~\ref{fig:fit_isoc_boots_gauss_abc}, the red dots show again
the minimum-entropy best-fit potential for the original sample. Panel
(a) shows the results for $10^4$ bootstraps, with red contours
representing percentiles 39.3 and 86.4 (1-$\sigma$ and 2-$\sigma$
equivalent contours in 2D). Panel (b) shows results for fits of $10^4$
data sets generated assuming $10\%$ Gaussian uncertainties in each of
the 6D coordinates, i.e. by sampling from Gaussian error distributions
centered on the original values (as one might do with observational
data, and correlated uncertainties could be introduced through a
covariance matrix). These results are biased towards higher masses due
to particles in the high-energy tail of the Gaussians, which are
unbound in lower mass models that are thus rejected, an issue not
present in the bootstrap samples.

\begin{figure*}
\begin{center}
	\includegraphics[width=\textwidth]{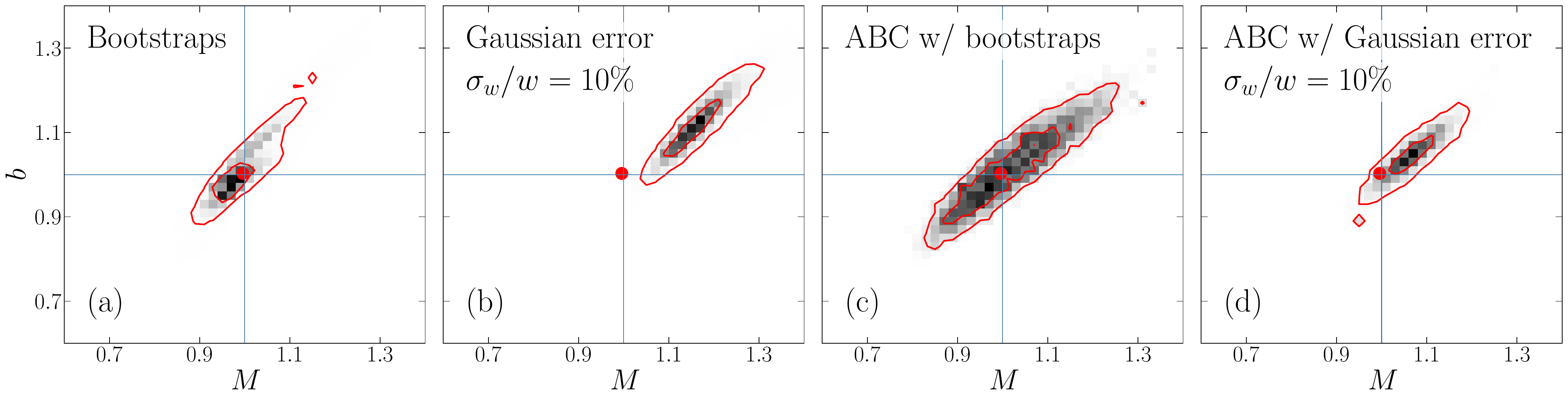}
        \caption{Fit results obtained for $10^4$ datasets generated as
          perturbations of an equilibrium sample with $N=10^4$
          particles. The minimum-entropy fit obtained with the
          original sample is shown as a red dot. Red lines are
          1-$\sigma$ and 2-$\sigma$ equivalent contours. Panel (a)
          shows fits for bootstrap samples; panel (b) shows fits
          re-sampling from Gaussian error distributions with relative
          uncertainties of $10\%$ in each coordinate; panel (c) [(d)]
          shows the ABC posteriors assuming flat priors, with data
          sets generated as in panel (a) [(b)].}
    \label{fig:fit_isoc_boots_gauss_abc}
\end{center}
\end{figure*}

Panels (a)-(b) represent the frequentist confidence contours on the
parameters, i.e. without explicitly introducing their prior
probabilities. For a Bayesian analysis, after finding a first estimate
of the best-fit potential whose DF is $f_0(\vec{J})$, we characterize
the potential's posterior probabilities. Without a \textit{bona fide}
likelihood, we cannot use traditional Markov Chain Monte Carlo
sampling, but we resort to a simulation-based inference
\citep[see][for a review]{Cranmer2020}. In particular, we perform an
Approximate Bayesian Computation \citep[ABC -- see][]{Beaumont2002,
  Sisson2018, Martin2021}, a sampling-rejection method that allows
sampling the posterior in problems where the likelihood is unknown or
intractable \citep[see e.g.][for applications in
cosmology]{Ishida2015, Hahn2017}.

We use the Python package pyABC \citep{pyabc2022}, which implements a
sequential Monte Carlo ABC \citep{Sisson2007}. We start drawing $\eta$
trial potentials from flat priors, with $0.1 \leq M \leq 5$ and
$0.1 \leq b \leq 5$. For each trial potential, we generate a new data
set, i.e. 6D coordinates $\vec{w}_i=(\vec{r}_i, \vec{v}_i)$ for
$i=1,\dots,N$. These new coordinates are generated either by
bootstrapping the original sample or by sampling from Gaussian error
distributions as explained above. For each trial potential, we
calculate actions for the associated $N$ coordinates. These actions
are then compared with the previously obtained actions of the original
sample in the best-fit minimum-entropy potential. For this comparison,
we estimate the KLD, Eq.~\eqref{eq:KLD} \citep[see e.g.][for its use
in ABC]{Jiang2018}. In practice, we calculate $\hat{S}_0$ once for the
best-fit potential, and $\hat{D}_\mathrm{KL}(f_0 || f)$ from the
cross-entropy $\hat{H}(f_0, f)$,
Eqs.~\eqref{eq:cross_entropy_def_2}-\eqref{eq:xi_est}, between actions
in the best-fit potential and those at each trial potential. We take
the median of all $\hat{D}_\mathrm{KL}(f_0 || f)$ to set a distance
threshold $\epsilon$ to be used in a next iteration. In each new
iteration, we draw new trial potentials from a probability
distribution built from weighted kernel density estimates of the
previously accepted potentials \citep[see][for
details]{Sisson2007}. Potentials are accepted if
$\hat{D}_\mathrm{KL}(f_0 || f) \leq \epsilon$ and each iteration
finishes when $\eta$ trial potentials are accepted.

The evolution of the sampling probability distribution towards the
parameters' posterior is driven by an adaptive decrease in the
distance threshold $\epsilon$ (calculated as the median of
$\hat{D}_\mathrm{KL}(f_0 || f)$ at each iteration), with a consequent
decrease in the samples acceptance rate. It is possible to show that
such sampling probability distribution iteratively converges to the
parameters' posterior -- see e.g. \cite{Beaumont2002, Sisson2007,
  Jiang2018}. In practice, convergence is assumed after the distance
threshold $\epsilon$ or the acceptance rate fall below a certain
value, or the changes in the posterior become negligible.

We iterate pyABC until the acceptance rate or the distance threshold
$\epsilon$ falls below $10^{-3}$, requiring $\eta = 10^4$ accepted
potentials in each iteration. Thus, the final iterations are slower
due to the high number of rejected potentials. The KLD is strictly
non-negative and in order to avoid negative KLD estimates due to noise
for potentials near the best fit, we actually take
$\max(\hat{D}_\mathrm{KL}(f_0 || f),10^{-6})$ as distance metric.

Fig~\ref{fig:fit_isoc_boots_gauss_abc} shows the ABC results with data
generated via bootstraps ({panel (c)) and by sampling from Gaussian
  error distributions with relative errors of $10\%$ for each
  coordinate (panel (d)). Note that the first estimate of the best fit
  potential (red dots) was obtained with the single original
  sample. For the ABC with bootstraps, pyABC runs up to iteration 9,
  when the distance threshold quickly drops to $\epsilon =
  10^{-6}$. This is due to our strategy to avoid negative KLD
  estimates mentioned above. The larger contours in comparison to
  panel (a) are due to this premature truncation of the procedure,
  indicating that in this case the KLD estimates are not precise
  enough to guarantee positive values for more iterations and better
  convergence. The true parameters are recovered with $\sim 3\%$
  errors and $\sim 10-12\%$ statistical uncertainties. Thus these
  contours are conservative estimates of the true ones, and
  encapsulate modeling uncertainties due to lack of better precision
  in the KLD estimates.

  For the ABC with $10\%$ Gaussian errors, pyABC runs up to iteration
  19, with the acceptance rate steadily declining to
  $\approx 10^{-3}$. In this case, we have $\sim 5-7\%$ errors and
  $\sim 6\%$ statistical uncertainties. This $\sim 5-7\%$ bias is
  still a manifestation of the Gaussian error distribution producing
  high-energy particles which exclude low-mass potentials, as seen in
  panel (b). However, since now the distance metric is anchored in the
  best-fit potential with DF $f_0(\vec{J})$ (not affected by the
  Gaussian re-sampling), this problem is alleviated and the bias
  reduced in respect to panel (b).

\subsection{Fitting an axisymmetric potential}
\label{sec:fit_mocks}

We now use a halo-like sample to fit an axisymmetric modified version
of the DM halo potential of \cite{McMillan2017}, where we introduce a
flattening parameter $q$, i.e. the ratio between the minor and major
axes. The potential is that associated with the density profile
\begin{equation}
    \frac{\rho_{DM}(\tilde{r})}{\rho_{0}} = \left(\frac{\tilde{r}}{r_s}\right)^{-\gamma}\left[ 1 + \left( \frac{\tilde{r}}{r_s}\right)\right]^{\gamma-3} \exp{\left[-\left(\frac{\tilde{r}}{400 \mathrm{kpc}}\right)^{6}\right]},
\end{equation}
where $\rho_{0} = 8.53702\times 10^6 M_\odot/\mathrm{kpc}^3$,
${r_s = 19.5725\kpc}$, $\gamma=1$,
$\tilde{r}=\sqrt{x^2 + y^2 + (z/q)^2}$ and $q=0.7$. The exponential
term is just a cutoff to assure a finite mass and to avoid numerical
problems. In principle, this DM halo potential could be added to all
the other components of the \cite{McMillan2017} potential (such as the
thin and thick discs), even if we only fit the parameters of the
former. However, in this case the inner potential would be dominated
by the baryonic components and the number of star-particles of our
tracer sample (described below) in the outer regions would not be
large enough to constrain the DM halo parameters. Therefore in what
follows we use the DM halo potential only.

We use \textsc{Agama} to generate a spherical stellar halo sample of
$N=10^4$ particles (the tracers), with a broken power-law density
profile given by
\begin{equation}
    \rho_h(r) \propto \left(\frac{r}{r_h}\right)^{-2.5}\left[ 1 + \left( \frac{r}{r_h}\right)\right]^{-0.5}\exp{\left[-\left(\frac{r}{300 \mathrm{kpc}}\right)^{3}\right]},
    \label{eq:rho_tracer}
\end{equation}
where $r_h = 25$ kpc, and the exponential term is again a cutoff at
large radii to avoid numerical problems. We set the velocities such
that this sample is stationary in our axisymmetric
potential. Specifically, we first create a sphericalized version of
the potential and initialize the isotropic DF using the Eddington
inversion formula, then express this DF as a function of actions,
embed it in the flattened potential, and sample positions and
velocities of stars from the resulting system. This procedure is
equivalent to adiabatically deforming the potential from the initial
(spherical) to the final (non-spherical) shape.

With this sample, assumed to be described by an unknown DF
$f(\vec{J})$, we fit the potential parameters $\rho_{0}$, $q$,
$\gamma$, and $r_s$. We estimate the actions
$\vec{J} = (J_r, J_\varphi, J_z)$ in each trial potential through the
Stackel fudge \citep{Binney2012} using \textsc{Agama}. As in
Sec.~\ref{sec:fit_isoc}, we first identify the best-fit potential
minimizing $\hat{S}_{\vec{J}}$, Eq.~\eqref{eq:S_J_est}, starting in a
grid of parameter values. We then use the actions in the globally
best-fit potential as the ``observed data'' described by an unknown DF
$f_0(\vec{J})$ to characterize the parameters' posterior in the ABC.

Once more, we run pyABC accepting $\eta = 10^4$ models in each
iteration, generating a new size-$N$ data sample for each
potential. New data sets are generated by sampling from Gaussian error
distributions of width $\sigma_{w_i}/|w_i|=10\%$ for
$i=1,\dots,6$. Fig.~\ref{fig:abc_dm_halo_10k_k10_err_0.01} shows the
resulting corner plot. Once more, we run pyABC until the acceptance
rate or the distance threshold falls below $10^{-3}$, by which time
the true parameters are well recovered. This suggests that this is a
reasonable choice when dealing with observed data, where the true
answer is unknown. In particular, the flattening parameter is
recovered with relative uncertainty $\sim 5\%$.

Fig.~\ref{fig:abc_dm_halo_10k_k10_err_0.2} shows a similar plot,
obtained with Gaussian error distributions with
$\sigma_{w_i}/|w_i|=20\%$ for each coordinate. We clearly see the
worsening of the fit compared to
Fig.~\ref{fig:abc_dm_halo_10k_k10_err_0.01}, but the true parameters
are still recovered reasonably well. In this case, the flattening
parameter is recovered with uncertainty $\sim 10\%$.

\begin{figure*}
    \centering
    \includegraphics[width=0.65\textwidth]{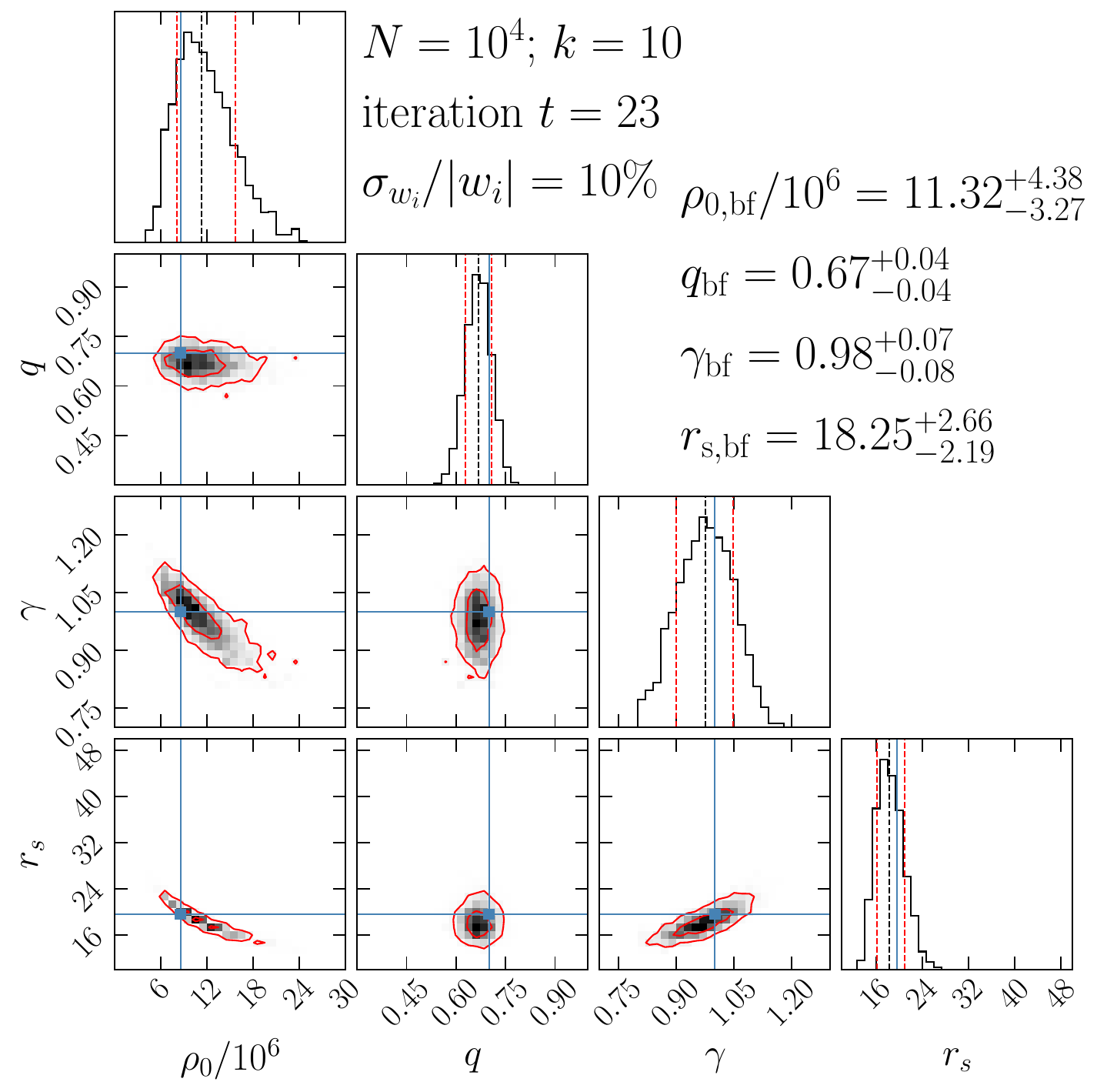}
    \caption{ABC results for $N=10^4$ particles that phase-mixed in a
      flattened axisymmetric potential ($q = 0.7$). 6D coordinates are
      assumed to have Gaussian error distributions with relative
      uncertainties $\sigma_{w_i}/|w_i| = 10\%$, for $i=1\dots 6$. In
      red, the 1-$\sigma$ and 2-$\sigma$ equivalent contours. Vertical
      dashed lines show the 16-th, 50-th and 84-th percentiles. The
      true parameters (blue lines/dots) are well recovered. In
      particular, for the flattening parameter $\sigma_q/q\sim 5\%$.}
    \label{fig:abc_dm_halo_10k_k10_err_0.01}
\end{figure*}

\begin{figure*}
    \centering
    \includegraphics[width=0.65\textwidth]{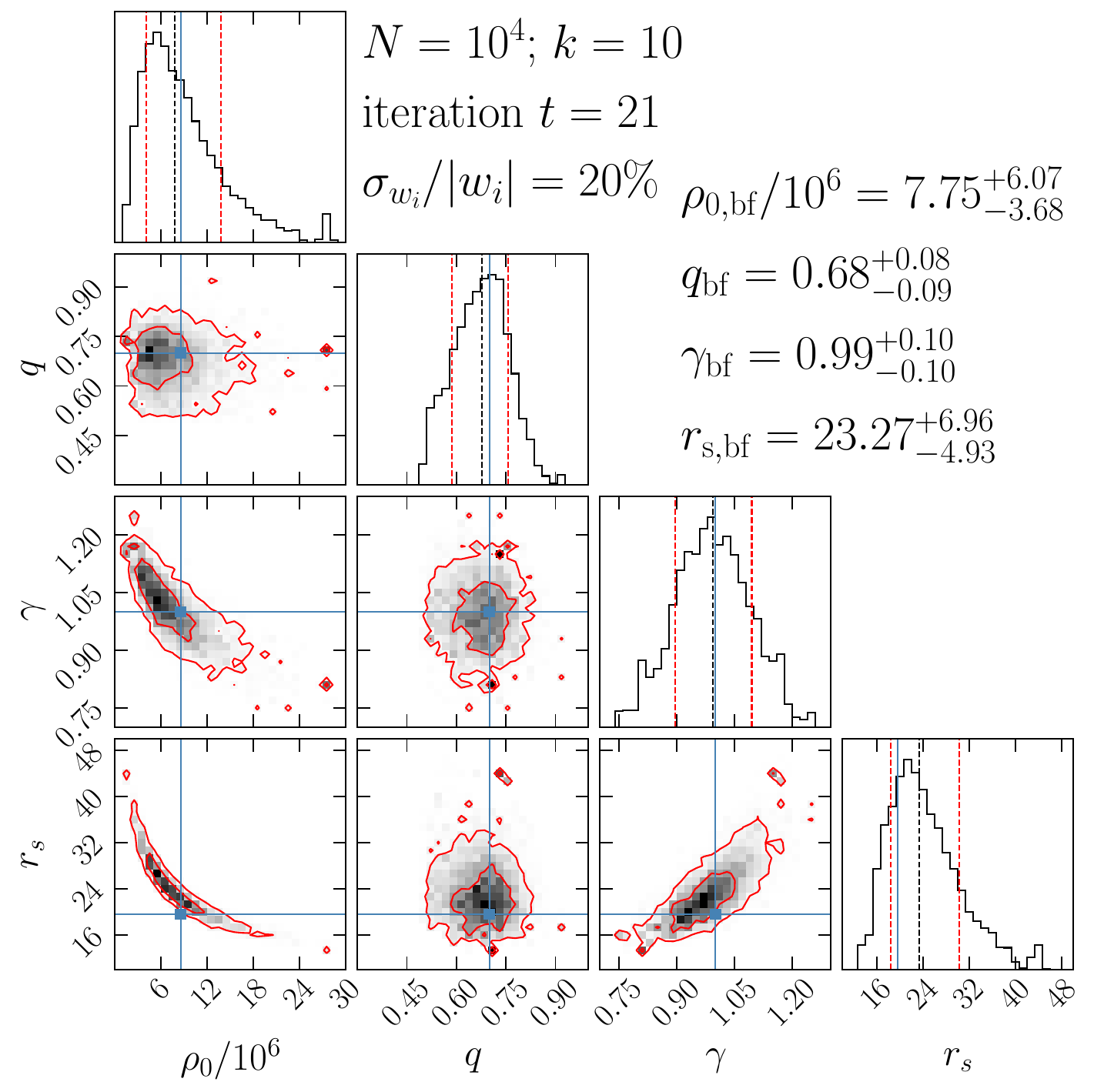}
    \caption{Similar to Fig.~\ref{fig:abc_dm_halo_10k_k10_err_0.01},
      but now assuming each coordinate to have Gaussian uncertainties
      $\sigma_{w_i}/|w_i| = 20\%$. We see the worsening of the fit,
      but the true parameters are still overall well recovered. In
      particular, $\sigma_q/q\sim 10\%$.}
    \label{fig:abc_dm_halo_10k_k10_err_0.2}
\end{figure*}

\section{Discussion}
\label{sec:discussion}

\subsection{Future improvements}
An ideal method to constrain a gravitational potential using the
kinematics of a stellar sample should:

\begin{enumerate}
\item allow constraints on general mass distributions, including
  general axisymmetric and triaxial systems;
    \item properly incorporate uncertainties and covariances in the
      data, providing not only best-fit values, but full probability
      distributions of the fit parameters;
    \item avoid making any assumptions regarding the DF besides the
      requirements of Jeans' theorem;
    \item be computationally efficient in order to handle samples with
      $\sim 10^4-10^6$ stars, typical of stellar halo samples, or
      stars within a globular cluster;
    \item properly consider the survey's footprint and selection
      function;
    \item handle incomplete information, e.g. samples missing
      line-of-sight velocities and/or distances.
    
\end{enumerate}

We demonstrated that our method already satisfies items 1--4. Although
we have not tested it for triaxial potentials, the only difficulty is
to efficiently estimate actions in such potentials. With these actions
at hand one can also investigate triaxial systems with this method.

In Sec.~\ref{sec:isochrone}, we discussed the bias and noise of the
k-NN entropy estimator used in this paper. Although this estimator is
good enough for most applications, our method would benefit from more
precise and accurate estimates -- see e.g. \cite{Lombardi2016,
  Berrett2019, Ao2023} for recent works with this aim.

In Sec.~\ref{sec:bias_correction}, we showed that the bias correction
proposed by \cite{Charzynska2015} effectively suppressed the bias in
the entropy estimates for self-consistent samples of the isochrone
model. This correction assumes the sample's support is a
parallelepiped defined by the extreme values of each coordinate. The
typical action-space of a self-consistent sample of an axisymmetric
potential has a shape close to a tetrahedron with two perpendicular
faces \citep[see Fig. 3.25 in][]{BT}. The reason why this simple
correction worked so well in the self-consistent isochrone sample is
probably that this tetrahedron support is not so different from the
assumed parallelepiped support for most stars. For non self-consistent
samples, and particularly for samples with sharp geometric cuts, the
actual support in action-space can be more complicated. For these
cases, it will be important to implement bias corrections for samples
with a general support.

In the DF-fitting method, where one assumes an analytical expression
for the DF, selection effects due to geometric cuts are taken into
account by the normalization factor
$A = {\int_\mathcal{V} f(\vec{w}|\vec{p}) \mathbb{S}(\vec{w})\dd^6
  \vec{w}}$, where $\mathcal{V}$ is the survey volume. This integral
can be very complicated and time consuming, and its limited numerical
accuracy is the main source of noise in these methods
\citep[][]{McMillan2013, Hattori2021}. In the minimum-entropy method
developed in this paper, we do not have an analytic DF, but the survey
footprint can be accounted by the fractional time each orbit spends in
it -- see Eq.~\eqref{eq:S_J_final}. This can be done either by
generating a number of angle variables uniformly distributed in
$[0 , 2\pi)$ for each star and checking how many pairs
$(\vec{\theta}, \vec{J})$ end up inside the footprint, or simply
integrating orbits and directly counting the fractional time inside
the footprint for each orbit. Other important improvements involve
handling samples with missing data and unbound stars, such as
hyper-velocity stars. In particular, we currently do not fit
potentials with even a single unbound star, which tends to bias the
estimates to higher masses if the original sample has high energy
stars in the correct potential (see
Fig.~\ref{fig:fit_isoc_boots_gauss_abc}). An improved version of the
method should handle (and penalize for) unbound stars to eliminate
this bias.

\subsection{Comparison with other methods}
In the DF-fitting method \citep[][]{McMillan2013, Hattori2021}, as the
traditional likelihood-based approach in general, the inference
process is facilitated by having a smooth function to be
maximized/minimized and by optimizing model evaluations. However, this
relies on the assumed DF correctly describing the data, which is hard
to guarantee in general, specially if deviations from equilibrium are
expected. The minimum-entropy method, in conjunction with the ABC
analysis, avoids that assumption while taking into account the
uncertainties in the data-generation process. In other words, the
possible bias introduced by assuming a DF in the DF-fitting method is,
in the minimum-entropy method, traded off for statistical
uncertainties that reflect our ignorance of the true DF.

\cite{Penarrubia2012} proposed to constrain the Galactic potential by
minimizing the entropy of the energy distribution of cold tidal
streams. Their method assumes a narrow energy distribution (in the
right potential), and that the probabilities for a star to be in a
certain position and to have a certain energy are independent. Under a
few approximations, they show that the entropy of the energy
distribution should be minimum at the true potential and, assuming a
Gaussian energy distribution, demonstrate that their method works for
spherical potentials. Analogously, \cite{Sanderson2015} proposed to
maximize the KLD between the action distribution and the product of
its marginal distributions of stellar streams. In other words, they
proposed to recover the true potential as the one maximizing the
correlations between the three actions. Their estimate of the KLD does
not require assuming a specific DF, but is performed on a fixed grid
in action space. When used to evaluate the odds of different models,
this requires rejecting points outside the grid of actions in the
best-fit model. \cite{Sanderson2015} apply their method to spherical
models, approximately recovering the true potential. \cite{Reino2021}
later constrained an axisymmetric Stackel potential using data on a
few streams, improving some aspects of this method. In particular,
they used EnLink \citep[][]{Sharma2009}, a metric-free density
estimator, to estimate the KLD -- but see also \cite{Reino2022} for an
erratum.

While these methods use stellar streams, assume the samples are
clustered in the space of integrals and minimize the entropy of the
integrals' pdf, our method targets smooth stellar populations, assumes
they are phase-mixed and minimizes the entropy of the full (6D) DF.

The orbital pdf method developed by \cite{Han2016} and its successor
emPDF \citep{Li2024} propose recovering the underlying potential
exploring Jeans' theorem but without specifying a DF, in a similar
vein as the minimum-entropy method developed here. Their methods are
currently restricted to spherical systems, but can be extended into
action space in a more general geometry. Their underlying general
principles and final expressions are similar to those we derive for
the spherical case, although based on different physical arguments and
developed independently. While \cite{Li2024} focus on estimating the
DF using Kernel Density Estimates, our approach uses well established
recipes to estimate the differential entropy of a sample via k-NN (but
a few other estimators can be used too).

Thus, in some sense, emPDF and the minimum-entropy method represent
two different views of the same general principles. Nonetheless, we
believe the general formalism developed in the current work
illuminates not only fundamental aspects of any method to constrain
mass distributions exploring Jeans' theorem, but also our picture of
the evolution of collisionless systems towards stationary states.

\subsection{Disequilibrium in the MW}
Complicating the application of the minimum-entropy method to the MW
is the kinematic perturbation from the Large Magellanic Cloud (LMC),
currently near a pericentric passage \citep[at a distance of
$\approx 50$ kpc,][]{Besla2007}. This perturbation is significant
enough to produce a reflex motion of the MW disc and its inner halo
($\lesssim 30$ kpc) towards the LMC past trajectory
\citep[][]{Garavito-Camargo2021, Petersen2021, Erkal2021}. Thus,
dynamical equilibrium cannot be assumed for the outer halo
($\gtrsim 30$ kpc). However, if one wants to probe the outer halo
still assuming dynamical equilibrium, a promising avenue is to try to
``undo'' or correct for the kinematical perturbation from the LMC
\citep[][]{Deason2021, Correa-Magnus2022}. On the other hand, for the
inner halo ($\lesssim 30$ kpc) the assumption of equilibrium still
seems reasonable.

\subsection{Time evolution versus fixed sample}

In Sec.~\ref{sec:entropy}, we introduced expressions for the entropy
of DFs that are functions of integrals of motion. On the one hand, one
can think of Eqs.~\eqref{eq:S_E_def}, \eqref{eq:S_EL_def} and
\eqref{eq:S_J_def} as the entropy the system would achieve if the same
sample is allowed to evolve in each trial potential until it
phase-mixes, with the original DF evolving to another DF that depends
only on integrals evaluated in that trial potential. In this case,
minimizing Eq.~\eqref{eq:S_E_def}, \eqref{eq:S_EL_def} or
\eqref{eq:S_J_def} corresponds to minimizing the future entropy, for
the sample will phase-mix if put in a wrong potential, increasing the
entropy - this generalizes the simpler reasoning of
\cite{Magorrian2014} for minimizing the entropy of an
``orbit-averaged'' DF. Interestingly, we do not need to wait for the
time evolution, since integrals are conserved and can therefore be
evaluated at the onset in each potential. The DF evolution is purely
driven by the remaining variables (e.g. angles), which evolve to a
uniform distribution in their respective supports.

On the other hand, in Appendix \ref{sec:appendix} we demonstrate that,
for a fixed equilibrium sample of a DF $f(\vec{r}, \vec{v})$,
\begin{equation}
    S_{\vec{I}} \geq S(f),
    \label{eq:S_I_geq_S_f}
\end{equation}
where $S(f)$ is the sample's invariant entropy and
\begin{equation}
    S_{\vec{I}} = -\int F(\vec{I})\ln \left[\frac{F(\vec{I})}{g(\vec{I})}\right]\dd \vec{I}
    \label{eq:S_I_general_text}
\end{equation}
is the general form of Eqs.~\eqref{eq:S_E_def}, \eqref{eq:S_EL_def}
and \eqref{eq:S_J_def}, with $\vec{I}$ being integrals, $g(\vec{I})$
the density of states and ${F(\vec{I}) = f(\vec{I})g(\vec{I})}$. In
this case, with no time evolution implied, the marginalization
defining the integrals' pdf $F(\vec{I})$, Eq.~\eqref{eq:def_pdf_I}, is
considered even when the remaining variables are not uniformly
distributed, i.e. considering the fixed sample as non-stationary in
each of the trial potentials. Furthermore, in Appendix
\ref{sec:appendix} we show that in the correct potential, where the
remaining variables are uniformly distributed on their supports,
$S_{\vec{I}} = S(f)$.

The two interpretations (future entropy and fixed-sample) require
minimizing the same quantity $S_{\vec{I}}$, showing that they are
equivalent. Thus, for potentials where stationary states are
synonymous with uniform distributions in the remaining variables (not
true in exceptional cases such as the harmonic oscillator), the
derivation of Eq.~\eqref{eq:S_I_geq_S_f} -- see Appendix
\ref{sec:appendix} -- represents a demonstration of the second law of
thermodynamics for collisionless gravitational systems in these
potentials, i.e. of the inevitable entropy increase for a sample
starting out of equilibrium, as illustrated in Figs. \ref{fig:hist_Jr}
and \ref{fig:S_isochrone_3M1}.

In contrast, it is traditionally assumed that the macroscopic
evolution of a collisionless system, i.e. in a smooth potential
$\phi$, is described by the Vlasov (or collisionless Boltzmann)
equation
\begin{equation}
    \frac{df}{dt} \equiv \frac{\partial f}{\partial t} + \vec{v}\cdot\frac{\partial f}{\partial \vec{r}} -\frac{\partial \phi }{\partial \vec{r}}\cdot\frac{\partial f}{\partial \vec{v}} = 0,
    \label{eq:vlasov}
\end{equation}
which implies entropy conservation. According to this view, the
aforementioned entropy increase would result from coarse-graining,
i.e. from losing information in fine scale phase-space structures
\citep[e.g.][]{LyndenBell1967, Tremaine1986, Dehnen2005, Levin2014,
  Barbieri2022, Banik2022}. Although the work of \cite{Dehnen2005} is
the closest to our interpretation, it still assumes that the
underlying evolution is described by Eq.~\eqref{eq:vlasov} (while
acknowledging that the extra fine phase-space structures introduced by
it are artificial), and that the evolution to a stationary state
requires coarse-graining.

Since coarse-graining is subjective, as it depends on the scale one
chooses to coarse-grain, the recovery of the gravitational potential
by minimizing the entropy of a sample in equilibrium would be
surprising if based on coarse-graining. Also surprising would be the
agreement of entropy estimates when using different sets of integrals,
since they involve distances and neighbors in very different
spaces. Additionally, the equivalence of
Eq.~\eqref{eq:S_I_general_text} with the expected future entropy after
phase-mixing in each trial potential might appear coincidental -- what
would be special about this coarse-grain scheme?

In line with \cite{BeS2019a, BeS2019b}, here we argue that these
entropy estimates recover all information that is available from a
finite-$N$ sample and thus do not operate a coarse-grain -- see
Sec.~\ref{sec:phase_mix_S_increase}. In contrast,
Eq.~\eqref{eq:vlasov} assumes the limit $N\rightarrow \infty$
\citep[][]{Dobrushin1979} and implies the development of indefinitely
fine phase-space structures, i.e. indefinitely large wave numbers $k$
in Fourier space. According to the Nyquist-Shannon theorem
\citep[][]{Nyquist1928, Shannon1949}, to a given size-$N$ sample in
$d$-dimensions one can only associate unique functions with maximum
wave number $k\lesssim N^{1/d}$. Functions with higher wave numbers
(sharper features) introduce information that is not contained in the
sample. This constrains the finest structures allowed for a DF
describing a real, i.e. finite-$N$, system \citep{BeS2019b}. Starting
out of equilibrium, the system starts developing fine-structures,
i.e. the maximum wave number increases, until hitting the
Nyquist-Shannon upper limit. After that, the system approaches a
steady state described by a DF that is a function of integrals only,
with the remaining variables uniformly distributed in their
domains. The time scale for this collisionless relaxation is
${\tau \lesssim 0.1 N^{1/6}\tau_\mathrm{cr}}$ \citep[][]{BeS2019a},
i.e. a few crossing times $\tau_\mathrm{cr}$ for typical stellar
samples. Thus, \emph{the system does not produce the extra fine
  phase-space structures predicted by Eq.~\eqref{eq:vlasov}}. For
recent related discussions in plasma physics, see \cite{Zhdankin2022,
  Zhdankin2023, Ewart2023, Nastac2024}.

In summary, real collisionless gravitational samples are finite-$N$
and, because of this, phase-mix towards stationary states described by
DFs depending only on integrals. Our method explores the objective
entropy increase associated with this process.

\section{Summary}
\label{sec:summary}
We have presented a method to constrain the gravitational potential
where a tracer sample is in dynamical equilibrium. It is based on the
idea that, if put in a different potential, this sample would
phase-mix, producing an entropy increase. The potential is then
recovered by minimizing the future entropy of the sample with respect
to the parameters of the potential. This entropy is estimated using
integrals of motion, and the parameters of the potential enter the fit
through these integrals.

We focused on actions, and demonstrated their advantages, including
possible constraints on the MW's DM halo shape. Investigation of this
particular problem will benefit from large spectroscopic surveys such
as the DESI-MWS \citep[][]{Cooper2023} in tandem with Gaia. The method
can be similarly applied to other integrals, such as energy and
angular momentum, e.g. in the study of spherical systems like globular
clusters -- see Appendix \ref{sec:appendix}. Finally, in Appendix
\ref{sec:max_entropy_angles} we discuss the possibility of recovering
a potential by maximizing the samples' entropy in angle-space,
concluding that this is not expected to work in general.

\section*{Acknowledgments}
We thank the anonymous referee for a careful reading and constructive
comments. LBeS thanks Wyn Evans, Josh Speagle, Chirag Modi, David
Hogg, Bernardo Modenesi, Sergey Koposov, Zhaozhou Li, Carrie Filion
and the stellar halos group at U. of Michigan for useful
discussions. MV \& LBeS acknowledge the support of NASA ATP award
80NSSC20K0509 and U.S. National Science Foundation AAG grant
AST-2009122, and MV acknowledges support from NASA ATP award
80NSSC24K0938. EV thanks Hans-Walter Rix and Kathryn Johnston for
valuable comments, and acknowledges support from an STFC Ernest
Rutherford fellowship (ST/X004066/1). KH is supported by JSPS KAKENHI
Grant Numbers JP24K07101, JP21K13965, and JP21H00053. WdSP is
supported by CNPq (309723/2020-5). LBeS \& KJD acknowledges support
from the Heising Simons Foundation grant \# 2022-3927. We respectfully
acknowledge that the U. of Arizona is on the land and territories of
Indigenous peoples. Today, Arizona is home to 22 federally recognized
tribes, with Tucson being home to the O’odham and the Yaqui. We
respect and honor the ancestral caretakers of the land, from time
immemorial until now, and into the future.

\software{numpy \citep{numpy},
          scipy \citep{scipy},
          Agama \citep{Vasiliev2019},
          pyABC \citep{pyabc2022},
          \texttt{tropygal} (this work).}


\appendix
\restartappendixnumbering

\section{Mathematical basis of the minimum-entropy method}
\label{sec:appendix}
For a DF separable in the space of angles-actions,
$f(\vec{\theta}, \vec{J}) = \mathcal{F}(\vec{\theta})F(\vec{J})$, the
entropy, Eq.~\eqref{eq:S_def}, is the sum of the respective
sub-spaces' entropies,
$S(f) = S(\mathcal{F}(\vec{\theta})) + S(F(\vec{J}))$. For stationary
states, the angle distribution is uniform,
$\mathcal{F}(\vec{\theta}) = (2\pi)^{-3}$, and thus
$S(\mathcal{F}(\vec{\theta}))$ is maximum, in the potential where the
sample is stationary. Since $S(f)$ is invariant for changes of
variables, i.e. for angle-actions evaluated in any potential,
$S(F(\vec{J}))$ is minimum in that potential.

Here we generalize this idea to non-separable DFs. In fact, one can
always separate pdfs in terms of conditional pdfs,
e.g.
$f(\vec{\theta}, \vec{J}) =
\mathcal{F}(\vec{\theta}|\vec{J})F(\vec{J})$, where
$\mathcal{F}(\vec{\theta}|\vec{J})$ is the conditional pdf of
$\vec{\theta}$, given $\vec{J}$. Thus, loosely speaking, we have
$S(f) = S(\mathcal{F}(\vec{\theta}|\vec{J})) + S(F(\vec{J}))$ and can
recover the potential by minimizing $S(F(\vec{J}))$, since
$S(\mathcal{F}(\vec{\theta}|\vec{J}))$ is maximum at any given action
in the right potential. Below we formalize this idea and generalize it
to other integrals.

Given the DF $f(\vec{w})$, where $\vec{w} \equiv (\vec{r}, \vec{v})$,
consider a random variable $X = (X_1,\dots,X_n)$, with
$X_i = X_i(\vec{w})$. Let $F_X$ be the pdf of $X$, i.e. $F_X$ is a
positive and normalized function on $\mathbb{R}^n$. The expectation
value of $X$ is
\begin{equation}
\mathbb{E}[X] = \int_{-\infty}^{\infty} f(\vec{w})X(\vec{w})\, \dd^6\vec{w} = \int_{-\infty}^{\infty} F_X(\vec{x})\vec{x}\, \dd^n\vec{x}.
\label{eq:expect_X}
\end{equation}
Let $\vec{I}=(I_{1},\ldots ,I_{m})$, with $m < 6$, be a second random
variable, with $I_i = I_i(\vec{w})$ -- we will later make $\vec{I}$ be
the integrals of motion, e.g.: for spherical and isotropic systems, we
set $I_1 = E$; for spherical and anisotropic ones, we set
$(I_1, I_2) = (E,L)$; for angle-action variables,
$(I_1, I_2, I_3) = \vec{J}$. In general, we require $\vec{I}$ to have
the following property:

\begin{quotation}
  1. There is a smooth function
  $\Psi :\mathbb{R}^{6}\rightarrow \mathbb{R}^{6}$, whose Jacobian
  matrix
\[
(J_{\Psi })_{ij}\doteq \frac{\partial \Psi _{i}}{\partial w_{j}}\text{ },%
\text{ \ \ }i,j=1,\ldots ,6\text{ }
\]%
is non-degenerate (i.e., its determinant is non-vanishing), such that
$I_{k}(\vec{w})=\Psi _{6-m+k}(\vec{w})$, $k=1,\ldots ,m$. In other
words, $\vec{I}$ corresponds to the last $m<6$ coordinates of some
change of variables $\Psi$ in 6D.
\end{quotation}

For random variables $\vec{I}$ with this property, we consider the
conditional expectation in the sense of a ``disintegration'' of $f$
with respect to $\vec{I}$,
$\mathcal{F}(\cdot|\vec{I})_{\vec{I}\in \mathbb{R}^{m}}$ -- for a
friendly, yet thorough, introduction to this topic, see
\cite{Chang1997}. This is a family of pdfs such that for each
$\vec{I}$, it gives the pdf $\mathcal{F}(\vec{z}|\vec{I})$ of the
remaining variables $\vec{z} \in \mathbb{R}^{6-m}$. This pdf is
properly normalized and is different from marginalizing over
$\vec{I}$, or from simply taking $f$ at fixed $\vec{I}$ values. Given
that the pdf of the new variables $(\vec{z}, \vec{I})$ is
$f\left( \Psi ^{-1}(\vec{z},\vec{I})\right) \cdot
|J_{\Psi^{-1}}(\vec{z},\vec{I})|$, and not
$f\left( \Psi ^{-1}(\vec{z},\vec{I})\right)$ alone, the conditional
probability with respect to $\vec{I}$ is explicitly given by:
\begin{equation}
\mathcal{F}(\vec{z}|\vec{I})=\frac{f\left( \Psi ^{-1}(\vec{z},\vec{I}%
)\right) |J_{\Psi ^{-1}}(\vec{z},\vec{I})|}{F(\vec{I})}\text{ },
\label{eq:cond_df_zI}
\end{equation}
where
\begin{equation}
F(\vec{I})=\int_{-\infty }^{\infty }f\left(\Psi^{-1}(\vec{z}^{\prime},\vec{I})\right)|J_{\Psi^{-1}
}(\vec{z}^{\prime},\vec{I})|\dd^{6-m}\vec{z}^{\prime} 
     \label{eq:def_pdf_I}
\end{equation}
is the pdf of the random variable $\vec{I}$, i.e., the marginalization
over the remaining variables $\vec{z}$. Moreover, as expected, $F$
only depends on $\vec{I}$, and not on the particular choice of the
transformation $\Psi:\mathbb{R}^{6}\rightarrow \mathbb{R}^{6}$ for the
remaining variables $\vec{z}$, since Eq.~\eqref{eq:def_pdf_I}
marginalizes over them. This elementary remark is important later on
in this appendix. Note that if $f = f(\vec{I})$, i.e. if it is uniform
in $\vec{z}$, Eq.~\eqref{eq:def_pdf_I} reduces to Eq.~\eqref{eq:F_E},
\eqref{eq:F_EL} or \eqref{eq:F_J} as particular cases.

With the change of variables $\vec{w} \rightarrow (\vec{z}, \vec{I})$
in Eq.~\eqref{eq:expect_X} we get:
\begin{equation*}
\mathbb{E}[X] =\int_{-\infty }^{\infty }f(\vec{w})X(\vec{w})\dd^6\vec{w}
=\int_{-\infty }^{\infty}X\left(\Psi^{-1}(\vec{z},\vec{I})\right)f\left(\Psi^{-1}(\vec{z},\vec{I})\right)|J_{\Psi}(\vec{I}, \vec{z})|\dd^{6-m}\vec{z}\textrm{ }\dd^{m}\vec{I},
\end{equation*}
and from Eq.~\eqref{eq:cond_df_zI} results that
\begin{eqnarray*}
\mathbb{E}[X] &=&\int_{-\infty}^{\infty}X\left(\Psi ^{-1}(\vec{z},\vec{I})\right) \mathcal{F}(\vec{z}|\vec{I}) F(\vec{I})  
\dd^{6-m}\vec{z}\textrm{ }\dd^{m}\vec{I} \\
&=&\int_{-\infty }^{\infty }F(\vec{I}) \Bigg(\int_{-\infty }^{\infty}\mathcal{F}(\vec{z}|\vec{I})X\left(\Psi^{-1}(\vec{z},\vec{I})\right)\dd^{6-m}\vec{z}\Bigg) \dd^{m}\vec{I}\textrm{}.
\end{eqnarray*}
In fact, the last equality is the formal \textit{definition} of
$\mathcal{F}(\cdot|\vec{I})_{\vec{I}\in \mathbb{R}^{m}}$ being the
disintegration of the DF with respect to the random variable
$\vec{I}$. Making $X = -\ln f$ and using Eq.~\eqref{eq:cond_df_zI}, we
get
\begin{eqnarray*}
S(f) =\mathbb{E}[-\ln f]&=&\int_{-\infty }^{\infty }F(\vec{I})\Bigg\{
\int_{-\infty }^{\infty }\mathcal{F}(\vec{z}|\vec{I})\cdot \left[ -\ln f(\Psi ^{-1}(\vec{z},\vec{I}))\right]
\dd^{6-m}\vec{z}\Bigg\} \dd^{m}\vec{I}\textrm{ } \\
&=&\int_{-\infty }^{\infty }F(\vec{I})\Bigg\{ \int_{-\infty }^{\infty}\mathcal{F}(\vec{z}|\vec{I})\cdot \left[ -\ln \left( \frac{F(\vec{I})}{|J_{\Psi^{-1}}(\vec{I}, \vec{z})|} \mathcal{F}(\vec{z}|\vec{I})\right) \right] \dd^{6-m}\vec{z}\Bigg\} \dd^{m}\vec{I}\textrm{ }.
\end{eqnarray*}
Writing the logarithm of the product as the sum of logarithms, we get
\begin{eqnarray*}
S(f) &=& \int_{-\infty }^{\infty}F(\vec{I})\Bigg\{ \int_{-\infty }^{\infty}  \mathcal{F}(\vec{z}|\vec{I}) \ln |J_{\Psi^{-1}}(\vec{I}, \vec{z})| \dd^{6-m}\vec{z}\Bigg\} \dd^{m}\vec{I}-\int_{-\infty }^{\infty}F(\vec{I})\ln F(\vec{I}) \dd^{m}\vec{I} \\
&&-\int_{-\infty }^{\infty }F(\vec{I})\Bigg\{\int_{-\infty }^{\infty}\mathcal{F}(\vec{z}|\vec{I})\ln \mathcal{F}(\vec{z}|\vec{I}) \dd^{6-m}\vec{z}\Bigg\}
\dd^{m}\vec{I}\textrm{ },
\end{eqnarray*}
where for the second term in the right-hand side we used the fact that
$\int_{-\infty }^{\infty}\mathcal{F}(\vec{z}|\vec{I})
\dd^{6-m}\vec{z}$ = 1. Hence,
\begin{equation}
S(f)=\mathbb{E}\left[\ln |J_{\Psi^{-1}}|\right]+S(F(\vec{I}))+\mathbb{E}_{\vec{I}}\left[ S(\mathcal{F}(\cdot | \vec{I}))\right] \textrm{ }, 
\label{eq:entropy:desint}
\end{equation}
where
$\mathbb{E}_{\vec{I}}\left[ S(\mathcal{F}(\cdot |\vec{I}))\right]$
denotes the expectation of the entropy
\begin{equation}
    S(\mathcal{F}(\cdot |\vec{I}))\doteq -\int_{-\infty }^{\infty }\mathcal{F}(\vec{z}|\vec{I})\ln
\left[ \mathcal{F}(\vec{z}|\vec{I})\right] d^{6-m}\vec{z}
\label{eq:entropy_conditional_df}
\end{equation}
of the conditional pdfs. Note that these entropies define a random
variable that only depends on $\vec{I}$. In particular, if $\Psi$ is a
canonical transformation ($|J_{\Psi^{-1}}(\vec{I}, \vec{z})|=1$), from
Eq.~\eqref{eq:entropy:desint} we have
$S(f)=S(F(\vec{I}))+\mathbb{E}_{\vec{I}}\left[
  S(\mathcal{F}(\cdot|\vec{I}))\right]$.

Eq.~\eqref{eq:entropy:desint} is the main general result of this
appendix. We show below that it justifies our minimum-entropy method
for fitting galactic potentials. With this aim, it is convenient to
make the following additional assumption on the variable
transformation $\Psi$, and afterwards we show how it can be removed:

\begin{quotation}
  2. The Jacobian determinant $|J_{\Psi^{-1}}(\vec{I},\vec{z})|$ only
  depends on $\vec{I}$.
\end{quotation}
In fact, given a partial transformation
$\tilde{\Psi} :\mathbb{R}^{6-m}\rightarrow \mathbb{R}^{6-m}$, it is
common to find a point transformation for the remaining variables such
that the total new transformation
$\Psi:\mathbb{R}^{6}\rightarrow \mathbb{R}^{6}$ is even
\emph{canonical}, i.e. $|J_{\Psi^{-1}}(\vec{I},\vec{z})| = 1$.

Suppose that, for all $\vec{I}\in \mathbb{R}^{m}$, the maximum allowed
support of the pdfs $\mathcal{F}(\cdot |\vec{I})$ is some bounded
region $\Omega _{z}(\vec{I})$ of $\mathbb{R}^{6-m}$. The region
$\Omega _{z}(\vec{I})$ encodes the set of coordinates
$\vec{z}\in \mathbb{R}^{6-m}$ corresponding to particles that, at
fixed $\vec{I}$, are \textit{not} forbidden to appear in the sample,
e.g. for being unbound or for its coordinates lying outside the survey
footprint.

If the coordinates $\vec{I}\in \mathbb{R}^{m}$ are constants of
motion, one expects that the original DF $f(\vec{w})$ is stationary
(phase-mixed), or, more generally, a cut of some stationary DF if, and
only if, $f(\Psi ^{-1}(\vec{z},\vec{I}))$ is constant for $\vec{z}$
within the maximum allowed support $\Omega _{z}(\vec{I})$, at any
fixed $\vec{I}\in $ $\mathbb{R}^{m}$. Thus, here we tacitly use this
property of the DF as equivalent to its stationarity. If condition
2. above is fulfilled then from Eq.~\eqref{eq:cond_df_zI}, for any
fixed $\vec{I}\in \mathbb{R}^{m}$, as a function of $\vec{z}$ the
conditional pdf $\mathcal{F}(\vec{z}|\vec{I})$ is proportional to
$f(\Psi ^{-1}(\vec{z},\vec{I}))$. Thus, one can detect that the DF
$f(\vec{w})$ is stationary, or a cut of a stationary DF, by showing
that the \emph{conditional} pdf $\mathcal{F}(\vec{z}|\vec{I})$ is
constant for $\vec{z}$ in $\Omega_{z}(\vec{I})$, at any fixed
$\vec{I}\in $ $\mathbb{R}^{m}$. We show now that this is equivalent to
our minimum-entropy principle.

A pdf supported on a fixed bounded region of $\mathbb{R}^{6-m}$ is
uniform if, and only if, it has maximal entropy. In this case,
$\mathcal{F}(\vec{z} |\vec{I}) = 1/V_{z}(\vec{I})$, where
$V_{z}(\vec{I})$ is the volume of the maximum allowed support
$\Omega_{z}(\vec{I})$ of $\mathcal{F}(\vec{z} |\vec{I})$, and from
Eq.~\eqref{eq:entropy_conditional_df},
\[
S(\mathcal{F}(\cdot |\vec{I}))=-\ln [1/V_{z}(\vec{I})]= \ln V_{z}(\vec{I})\text{ }. 
\]%
Thus, given a fixed DF $f$ on $\mathbb{R}^{6}$, the expected value
$\mathbb{E}_{\vec{I}}\left[ S(\mathcal{F}(\cdot |\vec{I}))\right] $ in
Eq.~\eqref{eq:entropy:desint} is bounded from above by
\[
\mathbb{E}_{\vec{I}}\left[ \ln V_{z}(\vec{I})\right] =\int_{-\infty
}^{\infty }F(\vec{I})\ln V_{z}(\vec{I})d^{m}\vec{I}\text{ }
\]%
and
$\mathbb{E}_{\vec{I}}\left[ S(\mathcal{F}(\cdot |\vec{I}))\right] $
reaches this value when the $\mathcal{F}(\cdot |\vec{I})$ are uniform
in their maximum allowed supports. Thus,
\begin{equation*}
\mathbb{E}_{\vec{I}}\left[ \ln V_{z}(\vec{I})\right] -\mathbb{E}_{\vec{I}}\left[ S(\mathcal{F}(\cdot |\vec{I}))\right]  \geq 0.
\end{equation*}
Using Eq.~\eqref{eq:entropy:desint},
\begin{equation*}
\begin{aligned}[b]
\mathbb{E}_{\vec{I}}\left[ \ln V_{z}(\vec{I})\right] -S(f)+ \mathbb{E}[\ln |J_{\Psi^{-1}}|]+S(F(\vec{I})) &\geq 0 \\
\mathbb{E}\left[ \ln \tilde{V}\right] +S(F(\vec{I})) &\geq S(f) \text{ },
\end{aligned}
\end{equation*}%
where 
$\tilde{V}(\vec{z},\vec{I})\doteq
|J_{\Psi^{-1}}(\vec{z},\vec{I})|V_{z}(\vec{I})$.

Therefore, by construction, the quantity 
$\mathbb{E}\left[ \ln \tilde{V}\right] +S(F(\vec{I}))$
is bounded from below by $S(f)$ and reaches this value if, and only
if, all $\mathcal{F}(\cdot |\vec{I})$ are uniform in their maximum
allowed supports. By assumption 2., we have:
\[
\mathbb{E}\left[ \ln \tilde{V}\right]  = \mathbb{E}_{\vec{I}}\left[ \ln g(\vec{I})\right], 
\]%
where $g(\vec{I}) = |J_{\Psi^{-1}}(\vec{I})|V_{z}(\vec{I})$ is the
``density of states'' at $\vec{I}\in \mathbb{R}^{m}$. Similar to
$F(\vec{I})$, the density of states $g(\vec{I})$ is independent of the
particular transformation $\Psi$ in respect to the remaining variables
$\vec{z}$. In fact, we can assume that there is a partial
transformation $\tilde{\Psi}$ over the $\vec{z}$ coordinates such that
the total transformation has
$|J_{\Psi^{-1}}(\vec{z}, \vec{I})| = |J_{\Psi^{-1}}(\vec{I})|$,
\begin{equation*}
g(\vec{I}) =|J_{\Psi ^{-1}}(\vec{I})|V_{z}(\vec{I}) =\int_{\Omega_{z}(\vec{I})}|J_{\Psi ^{-1}}(\vec{I})|d^{6-m}\vec{z}=\int_{\tilde{\Omega}_{z}(\vec{I})}|J_{\Psi ^{-1}}(\vec{I})|\frac{|J_{\tilde{\Psi}^{-1}}(\vec{z},\vec{I})|}{|J_{\Psi ^{-1}}(\vec{I})|}d^{6-m}\vec{z}
=\int_{\tilde{\Omega}_{z}(\vec{I})}|J_{\tilde{\Psi}^{-1}}(\vec{z},\vec{I})|d^{6-m}\vec{z}\text{ },
\end{equation*}
where $\tilde{\Omega}_{z}(\vec{I})$ is the maximum allowed support of
the remaining variables under the second transformation. Thus, the
density of states can be generalized for a transformation of
coordinates that does \emph{not} satisfy assumption 2. as
\begin{equation}
    g(\vec{I})\doteq \int_{\Omega _{z}(\vec{I})}|J_{\Psi ^{-1}}(\vec{z},\vec{I}%
)|d^{6-m}\vec{z}\text{ }.
\label{eq:dens_states}
\end{equation}
With this, we can finally relax condition 2. Hence, we proved, under
the assumption 1. only, that the quantity
\[
\mathbb{E}_{\vec{I}}\left[ \ln g(\vec{I})\right] +S(F(\vec{I}))
\]%
is bounded from below by $S(f)$ and reaches this value if, and only
if, $f(\Psi^{-1}(\vec{z},\vec{I}))$ is constant for $\vec{z}$ in
$\Omega _{z}(\vec{I})$.

Consider now a family $\Psi _{\vec{p}}$, $\vec{p}\in P$, of
transformations of coordinates in $\mathbb{R}^{6}$ satisfying
assumption 1., where $\vec{p}$ stands for generic parameters of the
potential. We think of $\Psi_{\vec{p}}$ as the set of transformations
leading to integrals of motion evaluated in all trial potentials. For
some fixed $m<6$, define $\vec{I}_{\vec{p}}$ by the last $m$
components of $\Psi _{\vec{p}}$, as above. Suppose, as before, that,
for all $\vec{p}\in P$ and $\vec{I}\in \mathbb{R}^{m}$, the maximum
allowed support of the pdfs $\mathcal{F}_{\vec{p}}(\cdot |\vec{I})$ is
some bounded region $\Omega _{z}(\vec{p},\vec{I})$ of
$\mathbb{R}^{6-m}$, which now can also depend on $\vec{p}$. If, for
some $\vec{p}_{0}\in P$, the DF $f$ is stationary (phase-mixed), or a
cut of a stationary DF, we can find this particular $\vec{p}_{0}$ by
minimizing with respect to $\vec{p}$ the quantity
\begin{equation}
    S_{\vec{I}} \doteq \mathbb{E}_{\vec{I}}\left[ \ln g_{\vec{p}}(\vec{I})\right] +S(F(\vec{I})) = -\int F(\vec{I})\ln \left[\frac{F(\vec{I})}{g_{_{\vec{p}}}(\vec{I})}\right]\dd^m \vec{I}.
    \label{eq:S_I_general}
\end{equation}
We emphasize that $S_{\vec{I}}$ incorporates the density of states
$g_{_{\vec{p}}}(\vec{I})$ and thus differs from $S(F(\vec{I}))$.  We
now show particular cases in terms of energy, angular momentum and
actions, making contact with Sec. \ref{sec:entropy}.

\subsection{Spherical and isotropic systems}
For spherically symmetric systems with isotropic velocities, we use
spherical coordinates, in terms of solid angles $\omega_r$ and
$\omega_v$, with Jacobian determinant
$\partial (\vec{r},\vec{v})/\partial (r,v,\omega_{r},\omega_{v})
=r^{2}v^{2}$.  For a given central potential $\phi_{_{\vec{p}}}(r)$,
the energy being ${E = v^2/2 + \phi_{_{\vec{p}}}(r)}$, the Jacobian
determinant for $(r,E)\rightarrow (r,v)$ is
$\partial (r,v)/\partial (r,E) = 1/\sqrt{2(E-\phi
  _{_{\vec{p}}}(r))}=1/v$. Thus,
\[
|J_{\Psi_{\vec{p}}}| = \frac{\partial (\vec{r},\vec{v})}{\partial (r,E,\varpi _{r},\varpi _{v})} =
\frac{\partial (\vec{r},\vec{v})}{\partial (r,v,\varpi _{r},\varpi _{v})} \frac{\partial (r,v)}{\partial (r,E)} = r^{2}\sqrt{2(E-\phi _{_{\vec{p}}}(r))}\text{ }.
\]
Let $r_{m}(E)$ be the maximum radius for a particle with energy
$E$. From Eq.~\eqref{eq:dens_states}, the density of states at fixed
$E$ is:
\begin{eqnarray*}
g_{_{\vec{p}}}(E) &=&\int_{(r,\omega_{r},\omega_{v})\in \Omega_{z}(E)}r^{2}\sqrt{2(E-\phi_{_{\vec{p}}}(r))}\dd r\dd\omega_{r}\dd\omega_{v} = (4\pi )^{2}\int_{0}^{r_{m}(E)}r^{2}\sqrt{2(E-\phi_{_{\vec{p}}}(r))}\dd r%
\text{ }.
\end{eqnarray*}
From ~\eqref{eq:S_I_general}, our minimum-entropy principle translates
into minimizing, with respect to the parameters $\vec{p}$,
\[
S_\mathrm{E} \doteq -\int F(E)\ln \left[\frac{F(E)}{g_{_{\vec{p}}}(E)}\right]\dd E\text{ },
\]
where $F(E)$ is the pdf for the energy -- c.f. Eq.~\eqref{eq:S_E_def}. Note that not only $g_{_{\vec{p}}}$, but also $F(E)$ depends on $\vec{p}$, via $\phi _{_{\vec{p}}}$.

\subsection{Spherical and anisotropic systems}
For a spherical system with anisotropic velocity distribution, we let
it depend on $v_{t}$ and $v_{r}$, the tangential and radial
velocities, respectively. The phase space coordinate
$(\vec{r},\vec{v})$ is a function of $r$, the solid angle
$\omega_{r}$, $v_{r}$ and $v_{t}$, as well as a planar angle
$\varphi_{v}$ referring to the tangent direction of the velocity,
i.e. we use cylindrical coordinates for the velocity $\vec{v}$, with
its vertical axis along $\vec{r}$. For this transformation of
coordinates we have
$\partial (\vec{r},\vec{v})/\partial (r,v_{r},v_{t},\omega
_{r},\varphi_{v})=r^{2}v_{t}$. With the angular momentum $L=rv_{t}$,
and $v^{2}=v_{t}^{2}+v_{r}^{2}$, we have
$v_{r}=\pm \sqrt{2(E-\phi _{_{\vec{p}}}(r))-L^{2}/r^{2}}$. Thus, the
Jacobian determinant of the transformation
$(E,L)\rightarrow (v_r,v_t)$ is
$\partial (v_{r},v_{t})/\partial (E,L) = \mp 1/\left[
  r\sqrt{2(E-\phi_{_{\vec{p}}}(r))-L^{2}/r^{2}}\right]$. Hence,
\[
|J_{\Psi_{\vec{p}}}| = \frac{\partial (\vec{r},\vec{v})}{\partial (r,E,L,\varpi _{r},\varphi _{v})} =\frac{\partial (\vec{r},\vec{v})}{\partial (r,v_{r},v_{t},\varpi
_{r},\varphi _{v})}\frac{\partial (v_{r},v_{t})}{\partial (E,L)}
=\mp \frac{L}{\sqrt{2(E-\phi _{_{\vec{p}}}(r))-L^{2}/r^{2}}}\text{ }.
\]
From Eq.~\eqref{eq:dens_states}, the density of states in this case is
$g_{_{\vec{p}}}(E,L)=8\pi^2 L T_r(E,L)$ --
c.f. Eqs.~\eqref{eq:g_EL}-\eqref{eq:T_r}. As before, from
~\eqref{eq:S_I_general} the minimum entropy principle refers to
minimizing
\[
S_\mathrm{EL} \doteq -\int F(E,L)\ln \left[\frac{F(E,L)}{g_{_{\vec{p}}}(E,L)}\right]\dd E\dd L\text{ },
\]%
where $F(E,L)$ is the joint pdf for the energy and angular momentum --
c.f. Eq.~\eqref{eq:S_EL_def}.

\subsection{Generic integrable potentials - action variables}
If $\Psi _{\vec{p}}$ is a canonical transformation,
$|J_{\Psi_{\vec{p}}^{-1}}|=1$. For instance, if $\Psi_{\vec{p}}$ refer
to action-angle variables, $\vec{I}_{\vec{p}}$ being actions, then, in
a full-sky survey, i.e. in the absence of any geometric cuts, from
Eq.~\eqref{eq:dens_states}:
\[
g_{\vec{p}}(\vec{I})=\tilde{V}_{\vec{p}}(\vec{I})=(2\pi )^{3}. 
\]%
From Eq.~\eqref{eq:S_I_general}, our minimum-entropy principle is
equivalent to minimizing
\[
S_{\vec{J}} \doteq -\int F(\vec{J}) \ln \left[\frac{F(\vec{J})}{(2\pi)^3} \right]\, \dd\vec{J},
\]
where $F(\vec{J})$ is the joint pdf for the actions --
c.f. Eq.~\eqref{eq:S_J_def}.

More generally, in the presence of geometrical cuts, 
\[
g_{\vec{p}}(\vec{I}) \rightarrow  g_{\vec{p}}(\vec{I}) A_{\vec{p}}(\vec{I}) ,
\]%
where the random variable $0 < A_{\vec{p}} \leq 1$ depends only on
$\vec{I}$ (integrals) and refers to the portion of the remaining
variables corresponding to stars lying within the survey footprint, at
fixed integral. Hence, in the presence of geometric cuts and when
using actions, our minimum entropy principle is equivalent to
minimizing
\begin{equation}
S_{\vec{J}} = -\int F(\vec{J}) \ln \left[\frac{F(\vec{J})}{(2\pi)^3A_{\vec{p}}(\vec{J})} \right]\, \dd\vec{J}. 
    \label{eq:S_J_final}
\end{equation}

\section{Could we maximize the entropy in angle-space?}
\label{sec:max_entropy_angles}
Since the angle distribution is uniform for a phase-mixed sample, one
might try to recover the potential by maximizing an entropy using
angles. In Sec.~\ref{sec:formalism}, we motivated our method by
connecting the maximum-likelihood principle with a minimum-entropy
one. This already suggests \emph{minimizing} an entropy using
integrals, as opposed to \emph{maximizing} one using the remaining
variables. Since these live in higher dimensions for the cases $f(E)$
and $f(E,L)$, it would not be helpful to use those variables, so for
this discussion we focus on angles and actions, both of which live in
$d=3$. We show why we do not expect a maximum-entropy in angle-space
to work.

As we demonstrate in Appendix \ref{sec:appendix}, writing the DF as
${f(\vec{\theta}, \vec{J}) =
  \mathcal{F}(\vec{\theta}|\vec{J})F(\vec{J})}$, we get
${S(f)=S\left(F(\vec{J})\right)+\mathbb{E}_{\vec{J}}\left[
    S(\mathcal{F}(\cdot | \vec{J}))\right]}$, where we have set
$|J_{\Psi^{-1}}| = 1$ in Eq.~\eqref{eq:entropy:desint}, $F(\vec{J})$
is the action's pdf and $S(\mathcal{F}(\cdot | \vec{J}))$ is the
entropy of the conditional pdfs --
Eq.~\eqref{eq:entropy_conditional_df}. The maximum value of this last
term, achieved in the potential where the sample is phase-mixed, is
$\mathbb{E}\left[\ln V_{\vec{\theta}}\right]$, where
${V_{\vec{\theta}}(\vec{J}) = (2\pi)^3 A(\vec{J})}$ is the volume of
the angles' support (density of states). In a full-sky survey,
$A(\vec{J})=1$, and $0 < A(\vec{J})< 1$ in the presence of geometric
cuts. This maximum value depends on the potential and needs to join
the optimization. Since we can calculate $V_{\vec{\theta}}(\vec{J})$
for each model, and $S(f)$ is invariant, the correct potential is
recovered if and only if $S_{\vec{J}}$ is minimum -- see
Eq. \eqref{eq:S_J_final}.

On the other hand, in trying to constrain the potential by maximizing
the entropy in angle-space, we would separate the DF as
${f(\vec{\theta}, \vec{J}) =
  \mathcal{F}(\vec{J}|\vec{\theta})F(\vec{\theta})}$, which implies
\begin{equation*}
    S(f)=S\left(F(\vec{\theta})\right)+\mathbb{E}_{\vec{\theta}}\left[ S(\mathcal{F}(\cdot | \vec{\theta}))\right],
\end{equation*}
where now $F(\vec{\theta})$ is the angles' pdf and
$S(\mathcal{F}(\cdot | \vec{\theta}))$ is the entropy of the
conditional pdfs. Although not easily justified, we could conjecture
that this last term is minimized in the potential where the sample is
phase-mixed, and that, as before, this depends on the potential and
should thus join the optimization. However, in this case we do not
know what this value should be and do not know which exact quantity to
maximize.

In principle, one might try simply maximizing the entropy of the
marginal pdf, $S\left(F(\vec{\theta})\right)$, but we show two
examples suggesting that this would fail. Let us consider an
admittedly artificial (1+1)D toy model with DF
\begin{equation*}
    f(\theta, J) =\frac{1}{\pi}\left[ \delta (\theta\leq \pi )\delta (J\leq 0)+\delta ( \theta>\pi )\delta ( J>0)\right],
\end{equation*}
where $\delta (\mathcal{P}) = 1$ when $\mathcal{P}$ is true, and
$\delta (\mathcal{P}) = 0$ otherwise, and $-1/2 \leq J \leq 1/2$. In
words, for negative actions, half of the angle maximum allowed support
$\Omega_{z}(J)= [0, 2\pi]$ is uniformly distributed, and for positive
actions the other half is. At fixed $J$, this DF is not constant as a
function of $\theta$ in $\Omega_{z}(J)$. Nevertheless, the marginal
$F(\theta) = \int f(\theta, J)\dd J$ is uniform, and $F(\theta)$ has
maximum entropy. Thus, maximizing $S\left(F(\theta)\right)$ generally
fails to reject non-stationary DFs.

As a second example, let $\Omega_{z}(J)=[0,4\pi |J|]$ for
$-1/2 \leq J \leq 1/2$ and $\Omega _{z}(J)=[0,2\pi ]$ when
$|J|>1/2$. This choice is to be understood as a ``toy model'' in the
presence of a geometric cut. Define the DF by
\[
f(\theta,J)=\frac{1}{\pi}\delta \left(\theta\in
\Omega_{z}(J)\right)\delta (|J|\leq 1/2)\text{ }.
\]%
This DF $f(\theta,J)$ is now uniform in angles. However, the marginal
pdf in this case is not uniform:
\[
F(\theta)= \int f(\theta, J)\dd J =\frac{1}{\pi}\left( 1-\frac{\theta}{2\pi }%
\right),
\]%
and thus its entropy is not maximum and maximizing
$S\left(F(\theta)\right)$ also generally fails to detect stationary
DFs.

\bibliography{refs}{}
\bibliographystyle{aasjournal}

\end{document}